\documentclass[pra,twocolumn,superscriptaddress,showpacs,floatfix]{revtex4-1}
\usepackage{graphicx,amssymb}
\usepackage[fleqn]{amsmath}
\usepackage{amsfonts}
\usepackage{dcolumn}% Align table columns on decimal point
\usepackage{bm}% bold math
\usepackage{braket}
\usepackage{multirow} 
\usepackage{MnSymbol}

\newcommand{\WignerSIXj}[6]
{
	\left\{
		\begin{array}{ccc}
			#1 & #2 & #3 \\
			#4 & #5 & #6
		\end{array}
	\right\}
}

\newcommand{\intt}
{
	\displaystyle \int
}

\newcommand{\hatvec}[1]
{
	\hat{ \vec{#1} }
}

\newcommand{\diffn}[3]
{
	\frac{ { \partial }^{ #3 } #1 }{ \partial { #2 }^{ #3 } }
}

\newcommand{\ddiff}[2]
{
	\frac{ d#1 }{ d#2 }
}

\newcommand{\Y}[2]
{
	{Y}_{ #2 }^{ #1 }
}

\newcommand{\hatY}[2]
{
	\hat{Y}_{ #2 }^{ #1 }
}

\newcommand{\sphRicatiBessel}[1]
{
	{S}_{ #1 }
}

\newcommand{\uvec}[1]
{
	\vec{u}_{ #1 }
}

\newcommand{\tensorop}[1]
{
	\hat{ \mathbf{#1} }
}

\newcommand{\units}[1]
{
	\, #1
}

\newcommand{\unitsText}[1]
{
	\, \text{#1}
}

\newcommand{\unitsMath}[1]
{
	\units{#1}
}

\begin{document}

\preprint{ }

\title{Description of the proton and neutron radiative capture reactions in the Gamow shell model }

\author{K. Fossez}
\affiliation{Grand Acc\'el\'erateur National d'Ions Lourds (GANIL), CEA/DSM - CNRS/IN2P3,
BP 55027, F-14076 Caen Cedex, France}

\author{N. Michel}
\affiliation{Grand Acc\'el\'erateur National d'Ions Lourds (GANIL), CEA/DSM - CNRS/IN2P3,
BP 55027, F-14076 Caen Cedex, France}

\author{M. P{\l}oszajczak}
\affiliation{Grand Acc\'el\'erateur National d'Ions Lourds (GANIL), CEA/DSM - CNRS/IN2P3,
BP 55027, F-14076 Caen Cedex, France}

\author{Y. Jaganathen}
\affiliation{Department of Physics \&
Astronomy, University of Tennessee, Knoxville, Tennessee 37996, USA}
\affiliation{NSCL/FRIB Laboratory,
Michigan State University, East Lansing, Michigan  48824, USA}

\author{R.M. Id Betan}
\affiliation{Physics Institute of Rosario (CONICET), Bv. 27 de Febrero 210 bis, S2000EZP Rosario, Argentina}
\affiliation{Department of Physics and Chemistry FCEIA(UNR), Av. Pellegrini 250, S2000BTP Rosario, Argentina}

\date{\today}

\begin{abstract}
	{ 
		We formulate the Gamow shell model (GSM) in coupled-channel (CC) representation for the description of proton/neutron radiative capture reactions and present the first application of this new formalism for the calculation of cross-sections in mirror reactions $^7$Be($p,\gamma$)$^8$B and $^7$Li($n,\gamma$)$^8$Li. The GSM-CC formalism is applied to a translationally-invariant Hamiltonian with an effective finite-range two-body interaction. Reactions channels are built by GSM wave functions for the ground state $3/2^-$ and the first excited state $1/2^-$ of $^7$Be/$^7$Li and the proton/neutron wave function expanded in different partial waves.
	}
\end{abstract}


\maketitle

\section{Introduction}
\label{sec0}

The description of nuclear structure and reactions in the unified theoretical framework is the long-standing challenge of nuclear theory. The attempts to reconcile the shell model (SM) with the reaction theory \cite{feshbach62_48,feshbach58_49} inspired the development of the continuum shell model (CSM) \cite{mahaux69_b16} which evolved into the unified of theory of nuclear structure and reactions \cite{mahaux69_b16,barz77_82,bennaceur00_44,volya06_94,volya05_470}. 

Structure of weakly bound states and resonances is different from the well-bound states. A comprehensive description of these systems  goes beyond standard configuration interaction model such as the SM and requires an open quantum system formulation of the many-body system. Such a generalization of the standard SM to describe well bound, weakly bound and unbound many-body states is provided by the Gamow shell model (GSM) \cite{michel02_8,michel03_10,michel09_2}. GSM offers the most general treatment of couplings between discrete and scattering states. The many-body states in GSM are given by the linear combination of Slater determinants defined in the Berggren ensemble of single particle states which consists of Gamow (resonant) states and the non-resonant continuum. 

In this formulation, GSM is the tool par excellence for studies of the structure of bound and unbound many-body states and their decays. For the description of reactions, the GSM has to be formulated in the CC representation. Recently, the GSM-CC has been applied for the calculation of excited states of $^{18}$Ne and $^{19}$Na, excitation function and the elastic/inelastic differential cross-sections in the  $^{18}$Ne$(p,p')$ reaction at different energies \cite{jaganathen12_551,jaganathen14_988}. In this work, we apply the GSM-CC formalism for the description of low-energy radiative capture reactions: ${ {}^{7}\text{Be} ( p , \gamma ) {}^{8}\text{B} }$ and ${ {}^{7}\text{Li} ( p , \gamma ) {}^{8}\text{Li} }$. In light nuclei, GSM-CC can be applied also for the description of nuclear reactions in the {\it ab initio} framework of the no-core GSM \cite{papadimitriou13_441} and to heavier projectiles like the $\alpha$-particle. 

The solution of solar neutrino problem is passing through an understanding of the $^7$Be($p,\gamma$)$^8$B proton capture reaction. $^8$B produced in the solar interior in this reaction, is the principal source of high energy neutrinos detected in solar neutrino experiments. At the solar energies ( $\sim$ 20 keV), this cross-section is too small to be directly measurable. For this reason, the theoretical analysis of this reaction is so important. On the other hand, whenever
the measurement is feasible ($E_{\rm CM} >$ 150 keV), the exact value of the capture cross section
depends: (i) on the normalization obtained indirectly from the $^7$Li($d, p$)$^8$Li cross section
and, (ii) on the model dependent extrapolation of measured values of the cross-section down
to the interesting domain of solar energies.

Proton radiative capture reaction on $^7$Be is of particular importance in astrophysics since it is involved in the pp-II and pp-III reaction chains. Indeed, the relative rates of the ${ {}^{7}\text{Be} ( {e}^{-} , { \nu }_{e} ) {}^{7}\text{Li} }$ reaction and the ${ {}^{7}\text{Be} ( p , \gamma ) {}^{8}\text{B} }$ reaction determine the pp-I/pp-II branching ratio, and thus the ratio of the neutrino fluxes coming from ${ {}^{7}\text{Be} }$ and ${ {}^{8}\text{B} }$ \cite{adelberger11_162}.
${ {}^{7}\text{Be} ( p , \gamma ) {}^{8}\text{B} }$ reaction has been studied experimentally by the direct proton capture \cite{filippone83_892,filippone83_887,hammache98_905,hammache01_906,junghans02_907,junghans03_889,baby03_900,baby03_908,baby04_901,baby04_909,junghans10_904} and the Coulomb dissociation of ${ {}^{8}\text{B} }$ \cite{baur86_886,schumann99_893,davids01_894,schumann03_895,schumann06_891}.
Theoretical approaches used to describe this reaction include the potential model \cite{kim87_158}, the R-matrix approach \cite{halderson06_903,barker95_885}, the shell model embedded in the continuum  (SMEC) \cite{bennaceur99_46}, the microscopic cluster model \cite{descouvemont04_890}, and the approach combining the resonating-group method and the no-core shell model \cite{navratil11_415}.

${ {}^{7}\text{Li} ( n , \gamma ) {}^{8}\text{Li} }$ reaction is the mirror reaction of ${ {}^{7}\text{Be} ( p , \gamma ) {}^{8}\text{B} }$.
$^7$Li($n,\gamma$)$^8$Li reaction cross section at very low energies provides the essential element of rapid process of primordial nucleosynthesis of nuclei with $A \geq 12$ in the inhomogeneous big-bang models \cite{applegate87_1109,fuller88_1110,malaney88_1107,terasawa89_1111}. Indeed, in
the inhomogeneous big-bang hypothesis, the main reaction chain leading to the
synthesis of heavy elements is \cite{malaney88_1107} $^1$H($n,\gamma$)$\rightarrow$$^2$H($n,\gamma$)$\rightarrow$$^3$H($d, n$)$\rightarrow$$^4$He($t,\gamma$)$\rightarrow$$^7$Li($n,\gamma$)$^8$Li, and then $^8$Li($\alpha, n$)$\rightarrow$$^{11}$B($n,\gamma$)$\rightarrow$$^{12}$B($\beta^-$)$\rightarrow$$^{12}$C($n,\gamma$)$\rightarrow$$^{13}$C$\rightarrow \dots$, etc., for heavier nuclei. In this sense, the reaction $^7$Li($n,\gamma$)$^8$Li is a key process to bridge the gap of mass $A = 8$ and to produce heavy elements. The reaction $^7$Li($n,\gamma$)$^8$Li has been studied experimentally \cite{wiescher89_916,heil98_896,nagai05_898,izsak13_914}. Theoretical studies of this reaction has been done using various potential models \cite{wang09_919,dubovichenko13_917}, the SMEC \cite{bennaceur99_46}, the microscopic cluster model \cite{descouvemont94_915} and the halo effective field theory approach \cite{rupak11_920}.

The paper is organized as follows. The chapter \ref{sec2} presents the general formalism of the GSM-CC approach. In Sec. \ref{sec2a}, we introduce the translationally-invariant GSM Hamiltonian in the cluster-orbital shell model (COSM) variables \cite{suzuki88_595}. The coupled-channel equations of the GSM-CC are presented in Sec. \ref{sec2b}. The channel states expansion in Berggren basis and the calculation of Hamiltonian matrix elements are discussed in Sects. \ref{sec2c} and \ref{subsubsec_CC_GSM_matrix_elts}, respectively. In Sec. \ref{subsubsec_CC_GSM_orthog}, we discuss how to orthogonalize the channel states, and Sec. \ref{subsubsec_CC_GSM_equiv_pot_meth} presents the method of solving the GSM-CC equations derived in this work.

The chapter \ref{sec_rad_cap_cross_sec} is devoted to the presentation of nucleon radiative capture formalism in the context of GSM-CC. In particular, the method of calculating matrix elements of the electromagnetic operators is explained in Sec. \ref{sec_rad_cap_method}, and the matrix elements itself are given in the appendix. 

Results of the GSM-CC calculations  are discussed in the chapters \ref{sec3} and \ref{sec4} for ${ {}^{7}\text{Be} ( p , \gamma ) {}^{8}\text{B} }$ and ${ {}^{7}\text{Li} ( p , \gamma ) {}^{8}\text{Li} }$ low-energy reactions, respectively.
Finally, main conclusions of this work are summarized in chapter \ref{sec5}.

\section{Coupled-channel formulation of the Gamow shell model} 
\label{sec2}

\subsection{Hamiltonian of the Gamow Shell Model}
\label{sec2a}

Center-of mass (CM) excitations in SM wave functions are removed  using Lawson method \cite{lawson80_b122,lipkin58_483,whitehead77_867}.
In the Gamow shell model (GSM), this method cannot be used because Berggren states are not eigenstates of the harmonic oscillator (HO) potential. To avoid spurious CM excitations in GSM wave functions, the GSM Hamiltonian is expressed in the intrinsic nucleon-core coordinates of the COSM \cite{suzuki88_595}:
\begin{equation}
	\hat{H} = \sum_{i = 1}^{ {N}_{ \text{val} } } \left( \frac{ \hatvec{p}_{i}^{2} }{ 2 { \mu }_{i} } + {U}_{c} ( \hat{r}_{i} ) \right) + \sum_{i < j}^{ {N}_{ \text{val} } } \left( V ( \hatvec{r}_{i} - \hatvec{r}_{j} ) + \frac{ {\hatvec{p}_{i}}{\cdot} {\hatvec{p}_{j} }}{ {M}_{c} } \right)
	\label{eq_GSM_Hamiltonian_COSM}
\end{equation}
where ${ {N}_{ \text{val} } }$ is the number of valence nucleons, ${ {M}_{c} }$ is the core mass, and:
$1/{ { \mu }_{i} } = 1/{ {M}_{c} } + 1/{ {m}_{i} }
\label{eq_GSM_COSM_reduced_mass}$,
is the reduced mass of the ${ i }$-th nucleon.
The single-particle potential ${ {U}_{c} ( \hat{r} ) }$ which describes the field of the core acting on each nucleon, is a sum of nuclear and Coulomb terms.  The nuclear term is given by a Woods-Saxon (WS) field with a spin-orbit term \cite{michel03_10}.
The Coulomb field is generated by a Gaussian density of $Z_c$ protons of the core \cite{michel03_10}. ${ V ( \hatvec{r}_{i} - \hatvec{r}_{j} ) }$ in \eqref{eq_GSM_Hamiltonian_COSM} is the two-body interaction which splits into nuclear and Coulomb parts.
As in the standard SM, adding and substracting a one-body mean-field ${ U ( \hat{r}_{i} ) }$ to the core Hamiltonian and the two-body interaction, respectively, allows to recast the GSM Hamiltonian in the form: 
\begin{equation}
{ \hat{H} = \hat{U}_{ \text{basis} } + \hat{T} + \hat{V}_{ \text{res} } }
\label{eq_H_GSM}
\end{equation} 
where the potential ${ \hat{U}_{ \text{basis} } }$ generates the s.p. basis, the kinetic term is written ${ \hat{T} }$ and the residual interaction is given by ${ \hat{V}_{ \text{res} }}$.
\begin{equation}
	\hat{U}_{ \text{basis} } = \sum_{ i = 1 }^{ {N}_{ \text{val} } } ( {U}_{c} ( \hat{r}_{i} ) + U ( \hat{r}_{i} ) )
	\label{eq_U_basis}
\end{equation}
\begin{equation}
	\hat{V}_{ \text{res} } = \sum_{i < j}^{ {N}_{ \text{val} } } \left( V ( \hatvec{r}_{i} - \hatvec{r}_{j} ) + \frac{ {\hatvec{p}_{i}}{\cdot} {\hatvec{p}_{j} }}{ {M}_{c} } \right) - \sum_{ i = 1 }^{ {N}_{ \text{val} } } U ( \hat{r}_{i} )
	\label{eq_H_res}
\end{equation}
In the present studies, we use the Furutani-Horiuchi-Tamagaki (FHT) finite-range two-body interaction \cite{furutani78_1012,furutani79_1013}:
\begin{equation}
	V ( \hatvec{r}_{i} - \hatvec{r}_{j} ) \equiv {V}_{ i j } = {V}_{ i j }^{ \text{C} } + {V}_{ i j }^{ \text{SO} } + {V}_{ i j }^{ \text{T} } + {V}_{ i j }^{ \text{Co} } \ .
	\label{eq_FHT_inter_0}
\end{equation}
The central potential ${ {V}_{ i j }^{ \text{C} } }$ is:
\begin{equation}
	{V}_{ i j }^{ \text{C} } = \sum_{ n = 1 }^{3} {V}_{ 0 , n }^{ \text{C} } {e}^{ - { \beta }_{n}^{ \text{C} } {r}^{2} } ( {W}_{n}^{ \text{C} } + {B}_{n}^{ \text{C} } {P}^{ \sigma } - {H}_{n}^{ \text{C} } {P}^{ \tau } - {M}_{n}^{ \text{C} } {P}^{ \sigma } {P}^{ \tau } ) \ ,
	\label{eq_FHT_inter_central_pot_0}
\end{equation}
where ${ r }$ is the distance between particles $i$ and $j$, ${ { \beta }_{n}^{ \text{C} } }$ is the range of gaussians, ${ {P}^{ \sigma } }$ and ${ {P}^{ \tau } }$ are the spin exchange and isospin exchange operators, respectively, and ${ {W}_{n}^{ \text{C} } }$, ${ {B}_{n}^{ \text{C} } }$, ${ {H}_{n}^{ \text{C} } }$ and ${ {M}_{n}^{ \text{C} } }$ are the exchange parameters.
The spin-orbit potential ${ {V}_{ i j }^{ \text{SO} } }$ writes:
\begin{equation}
	{V}_{ i j }^{ \text{SO} } = {\vec{L}} {\cdot} {\vec{S} }\sum_{ n = 1 }^{2} {V}_{ 0 , n }^{ \text{SO} } {e}^{ - { \beta }_{n}^{ \text{SO} } {r}^{2} } ( {W}_{n}^{ \text{SO} } - {H}_{n}^{ \text{SO} } {P}^{ \tau } )
	\label{eq_FHT_inter_so_pot_0}
\end{equation}
where ${ \vec{L} }$ is the relative orbital angular momentum between the two particles and ${ \vec{S} = \vec{s}_{i} + \vec{s}_{j} }$ where ${ \vec{s}_{ i } }$, ${ \vec{s}_{ j } }$ are the spins of particles ${ i }$, $j$.
The tensor potential ${ {V}_{ i j }^{ \text{T} } }$ writes:
\begin{equation}
	{V}_{ i j }^{ \text{T} } = {O}_{ \text{T} } \sum_{ n = 1 }^{3} {V}_{ 0 , n }^{ \text{T} } {e}^{ - { \beta }_{n}^{ \text{T} } {r}^{2} } ( {W}_{n}^{ \text{T} } - {H}_{n}^{ \text{T} } {P}^{ \tau } )
	\label{eq_FHT_inter_tensor_pot_0}
\end{equation}
where 
\begin{equation}
	{O}_{ \text{T} } = \left( \frac{ 3 ( {\vec{ \sigma }_{i}} {\cdot} {\vec{r}} ) ( {\vec{ \sigma }_{j}} {\cdot} {\vec{r}} ) }{ {r}^{2} } - {\vec{ \sigma }_{i}} {\cdot} {\vec{ \sigma }_{j}} \right) {r}^{2}
	\label{eq_FHT_inter_tensorial_op_0}
\end{equation}
and ${ \vec{ \sigma }_{ i } }$, ${ \vec{ \sigma }_{ j } }$ are the Pauli matrices. The Coulomb potential in the FHT interaction is standard. 

It is convenient to rewrite the FHT interaction using projection operators on singlet and triplet states of spin and isospin:
\begin{align}
	{V}^{ \text{C} } (r) &= {\cal V}_{ \text{tt} }^{ \text{C} } {f}_{ \text{tt} }^{ \text{C} } (r) { \pi }_{ \text{t} }^{ \sigma } { \pi }_{ \text{t} }^{ \tau } + {\cal V}_{ \text{ts} }^{ \text{C} } {f}_{ \text{ts} }^{ \text{C} } (r) { \pi }_{ \text{t} }^{ \sigma } { \pi }_{ \text{s} }^{ \tau } \nonumber \\
	&+ {\cal V}_{ \text{st} }^{ \text{C} } {f}_{ \text{st} }^{ \text{C} } (r) { \pi }_{ \text{s} }^{ \sigma } { \pi }_{ \text{t} }^{ \tau } + {\cal V}_{ \text{ss} }^{ \text{C} } {f}_{ \text{ss} }^{ \text{C} } (r) { \pi }_{ \text{s} }^{ \sigma } { \pi }_{ \text{s} }^{ \tau }
	\label{eq_FHT_inter_central_pot_1}
\end{align}
\begin{equation}
	{V}^{ \text{SO} } (r) = {\vec{L}}{\cdot} {\vec{S}} ( {\cal V}_{ \text{tt} }^{ \text{SO} } {f}_{ \text{tt} }^{ \text{SO} } (r) { \pi }_{ \text{t} }^{ \sigma } { \pi }_{ \text{t} }^{ \tau } + {\cal V}_{ \text{ts} }^{ \text{SO} } {f}_{ \text{ts} }^{ \text{SO} } (r) { \pi }_{ \text{t} }^{ \sigma } { \pi }_{ \text{t} }^{ \tau } )
	\label{eq_FHT_inter_so_pot_1}
\end{equation}
\begin{equation}
	{V}^{ \text{T} } (r) = {O}_{ \text{T} } ( {\cal V}_{ \text{tt} }^{ \text{T} } {f}_{ \text{tt} }^{ \text{T} } (r) { \pi }_{ \text{t} }^{ \sigma } { \pi }_{ \text{t} }^{ \tau } + {\cal V}_{ \text{ts} }^{ \text{T} } {f}_{ \text{ts} }^{ \text{T} } (r) { \pi }_{ \text{t} }^{ \sigma } { \pi }_{ \text{t} }^{ \tau } )
	\label{eq_FHT_inter_tensor_pot_1}
\end{equation}
where ${ { \pi }_{ \text{s} }^{ \sigma } }$, ${ { \pi }_{ \text{t} }^{ \sigma } }$ are the projection operators on singlet and triplet states of spin:
\begin{equation}
	{ \pi }_{ \text{s} }^{ \sigma } = \frac{1}{4} - {\vec{s}_{i}} {\cdot} {\vec{s}_{j}} \ ; \quad 
	{ \pi }_{ \text{t} }^{ \sigma } = \frac{3}{4} + {\vec{s}_{i}} {\cdot} {\vec{s}_{j}} \ ; \quad
	{P}^{ \sigma } = \frac{1}{2} + 2 {\vec{s}_{i}} {\cdot} {\vec{s}_{j}} = { \pi }_{ \text{t} }^{ \sigma } - { \pi }_{ \text{s} }^{ \sigma }
	\label{eq_FHT_inter_proj_op_singlet_triplet_spin}
\end{equation}
Eqs. \eqref{eq_FHT_inter_proj_op_singlet_triplet_spin} have the same form for projection operators ${ { \pi }_{ \text{s} }^{ \tau } }$ and ${ { \pi }_{ \text{t} }^{ \tau } }$ on singlet and triplet states of isospin.
Functions ${ {f}_{ \sigma \tau } (r) }$ in Eqs. \eqref{eq_FHT_inter_central_pot_1}-\eqref{eq_FHT_inter_tensor_pot_1}
depend on the parameters ${ {V}_{ 0 , n }^{\zeta} }$, ${ {W}_{n}^{\zeta} }$, ${ {B}_{n}^{\zeta} }$, ${ {H}_{n}^{\zeta} }$, ${ {M}_{n}^{\zeta} }$, $\beta_n^{\zeta}$, where $\zeta$ stands for superscripts 'C', 'SO' and 'T', which are given in Refs. \cite{furutani78_1012,furutani79_1013}.
In this work, we adjust the coupling constants ${ {\cal V}_{ a a' }^{ \zeta } }$ in Eqs.(\ref{eq_FHT_inter_central_pot_1}-\ref{eq_FHT_inter_tensor_pot_1}), where ${ a }$ and ${ a' }$ are indices of either singlet (s) or triplet (t) states.

\subsection{The coupled-channel equations}
\label{sec2b}

Nuclear reactions can be conveniently formulated in the CC representation of the Schr\"odinger equation. 
The first step to derive the GSM-CC equations is to expand the GSM eigenstates 
in the complete basis of channel states
${ \{ \ket{c} \} } \equiv { \{ \ket{ {c_{\rm proj}; c_{\rm targ}} }   \} }$ 
which contain information about the structure of the target and the projectile. Indices ${ {c}_{ \text{proj} } }$ and ${ {c}_{ \text{targ} } }$ denote the sets of quantum numbers associated with the projectile and the target, respectively. The nuclear reaction is then described by the relative motion of target and projectile nuclei and the channel parameters, like angular momenta of the target and the projectile and quantum numbers of the target internal excitations. In the following discussion, the heavy reaction participant is called a 'target' and the light one a 'projectile'. Obviously, the formulation of reaction theory in the GSM-CC approach does not depend on this arbitrary choice of labels.

The antisymmetric eigenstates of GSM-CC equations
\begin{equation}
	\hat{ \mathcal{A} }\ket{ \Psi } = \ket{ \Psi } = \sumint\limits_{c} \intt_{0}^{ \infty } dr \, {r}^{2} \braket{ r , c | \Psi } \ket{ r , c } \ ,
	\label{eq_as_eig_restricted_basis}
\end{equation}
where ${ \hat{ \mathcal{A} } }$ is the antisymmetrization operator, can be expanded using the channel basis states: ${ \ket{ r , c } = \hat{ \mathcal{A} } ( \ket{r} \otimes \ket{c} ) }$. 
In the above equation, ${ \braket{ r , c | \Psi } }$ are the antisymmetrized channel wave functions:
${ \Psi }_{c} (r) \equiv \braket{ r , c | \Psi } \equiv  {u}_{c} (r) r
\label{eq_as_channel_wfs}$.
Hence:
\begin{equation}
	\ket{ \Psi } = \sumint\limits_{c} \intt_{0}^{ \infty } dr \, {r}^{2} \frac{ {u}_{c} (r) }{r} \ket{ r , c }  \ .
	\label{eq_expan_eig_full_channel_basis}
\end{equation}

GSM-CC equations are obtained by inserting \eqref{eq_expan_eig_full_channel_basis} in the Schr\"odinger equation and then projecting this equation on a given channel basis state ${ \bra{ r' , c' } }$. One obtains:
\begin{equation}
	\sumint\limits_{c} \intt_{0}^{ \infty } dr \, {r}^{2} \left( {H}_{ c' , c } ( r' , r ) - E {N}_{ c' , c } ( r' , r ) \right) \frac{ {u}_{c} (r) }{r} = 0 
	\label{eq_CC_eqs_general}
\end{equation}
where:
\begin{equation}
	{H}_{ c' , c } ( r' , r ) = \braket{ r' , c' | \hat{H} | r , c }
	\label{eq_CC_H_ME_general}
\end{equation}
and
\begin{equation}
	{N}_{ c' , c } ( r' , r ) = \braket{ r' , c' | r , c }
	\label{eq_CC_N_ME_general}
\end{equation}
are the Hamiltonian matrix elements and the norm matrix elements in the channel representation, respectively.

\subsection{Channel states expansion in the Berggren basis}
\label{sec2c}

In the present studies, any target state ${ \ket{ {c}_{ \text{targ} } } }$ is an antisymmetrized state of ${ A - 1 }$ nucleons:
\begin{equation}
	\ket{ c_{\rm targ} } = \sum_{i} \braket{ \text{SD}_{i}^{ ( A - 1 ) } | c_{\rm targ }} \ket{ \text{SD}_{i}^{ ( A - 1 ) } } = \sum_{i} {a}_{ i , {c}_{ \text{targ} } } \ket{ \text{SD}_{i}^{ ( A - 1 ) } }
	\label{eq_targ_state_expansion_sd}
\end{equation}
Slater determinants ${ \ket{ \text{SD}_{i}^{ ( A - 1 ) } } }$ are built using a complete set of single-particle states of the Berggren ensemble which includes both resonant states and complex-energy scattering states.
Berggren ensemble is generated by the single-particle potential ${ \hat{U}_{ \text{basis} } }$ acting on the valence nucleons. This ensemble is also used to generate the states of the projectile:
\begin{equation}
	\ket{ { \phi }_{i;c_{\rm proj}} } = \hat{ \mathcal{A} } ( \ket{ { \phi }_{i}^{ \text{rad} } } \otimes \ket{c_{\rm proj} }) = \hat{ \mathcal{A} } ( \ket{ { \phi }_{i}^{ \text{rad} } } \otimes \ket{ l , s ; j , {m}_{j} } ) 
	\label{eq_berggren_state_separation_rad_ang}
\end{equation}
where $\ket{ { \phi }_{i}^{ \text{rad} }}$ and ${ \ket{c_{\rm proj} } }$ are radial and angular parts, respectively. In this expression,
${ l }$ is the orbital angular momentum of the nucleon, ${ s }$ its spin, ${ j }$ is the total angular momentum and ${ {m}_{j} }$ its projection. Hence, the basis state ${ \ket{ r , {c}_{ \text{proj} } } }$ of a projectile can be written as:
\begin{equation}
	\ket{ r , {c}_{ \text{proj} } } = \sum_{i} \frac{ {u}_{i} (r) }{r} \ket{ { \phi }_{i;c_{\rm proj} }}
	\label{eq_proj_state_berggren_expansion_2}
\end{equation}
where ${u}_{i} (r) / r  = { \braket{ { \phi }_{i}^{ \text{rad} } | r }}$.
Using Eq. \eqref{eq_proj_state_berggren_expansion_2}, one can write the channel basis states as:
\begin{equation}
	\ket{ r , c } = \sum_{i} \frac{ {u}_{i} (r) }{r} \ket{ { \phi }_{i}^{ \text{rad} } , c }
	\label{eq_CC_basis_state_expansion_Berggren_2}
\end{equation}
where ${ \ket{ { \phi }_{i}^{ \text{rad} } , c } = \hat{ \mathcal{A} } ( \ket{ { \phi }_{i}^{ \text{rad} } } \otimes \ket{c} ) }$.

\subsection{Hamiltonian matrix elements}
\label{subsubsec_CC_GSM_matrix_elts}

Matrix elements of the Hamiltonian ${ {H}_{ c' , c } ( r' , r ) }$ and the norm ${ {N}_{ c' , c } ( r' , r ) }$  can be derived using the expansion \eqref{eq_CC_basis_state_expansion_Berggren_2} which allows to treat the antisymmetry in the projectile-target system. In practice, only a finite number of Slater determinants contribute significantly to the target state and thus the antisymmetry between the low-energy target states and the high-energy projectile states can be neglected in most cases. The high-energy terms correspond to the channel basis states with high momentum ${ k }$ or high-${ i}$ indices ${ i > {i}_{ \text{max} } }$, where ${ {i}_{ \text{max} } }$ depends on the considered channel ${ c }$.
Hence, the expansion \eqref{eq_CC_basis_state_expansion_Berggren_2} can be splitted into low- and high-energy parts:
\begin{align}
	\ket{ r , c } &= \sum_{ i = 1 }^{ {i}_{ \text{max} } - 1 } \frac{ {u}_{i} (r) }{r} \ket{ { \phi }_{i}^{ \text{rad} } , c } + \sum_{ i = {i}_{ \text{max} } }^{N} \frac{ {u}_{i} (r) }{r} \ket{ { \phi }_{i}^{ \text{rad} } , c } \nonumber \\
	&\simeq \sum_{ i = 1 }^{ {i}_{ \text{max} } - 1 } \frac{ {u}_{i} (r) }{r} \ket{ { \phi }_{i}^{ \text{rad} } , c } + \sum_{ i = {i}_{ \text{max} } }^{N} \frac{ {u}_{i} (r) }{r} \ket{ { \phi }_{i}^{ \text{rad} } } \otimes \ket{c}
	\label{eq_CC_basis_state_exp_Berggren_split}
\end{align}
where ${ N }$ is the number of discretized continuum states and ${ {i}_{ \text{max} } }$ is the index from which the antisymmetry effects are neglected. Equivalently, Eq. \eqref{eq_CC_basis_state_exp_Berggren_split} can be written as:
\begin{align}
	\ket{ r , c } &= \sum_{ i = 1 }^{ {i}_{ \text{max} } - 1 } \frac{ {u}_{i} (r) }{r} \ket{ { \phi }_{i}^{ \text{rad} } , c } \nonumber \\
	&+ \ket{r} \otimes \ket{c} - \sum_{ i = 1 }^{ {i}_{ \text{max} } - 1 } \frac{ {u}_{i} (r) }{r} \ket{ { \phi }_{i;c_{\rm proj}} } \otimes \ket{ {c}_{ \text{targ} } }
	\label{xxx}
\end{align}
where $\ket{r} \otimes \ket{c}$ and $\ket{ { \phi }_{i;c_{\rm proj}} } \otimes \ket{ {c}_{ \text{targ} }}$ stand for non-antisymmetrized states. In this particular case (${ i \geq {i}_{ \text{max} } }$), the GSM Hamiltonian \eqref{eq_H_GSM}  
splits into ${ \hat{H}_{ \text{proj} } }$ and ${ \hat{H}_{ \text{targ} } }$ terms acting on projectile states ${ \ket{ { \phi }_{i;c_{\rm proj}} } }$ and target states ${ \ket{ {c}_{ \text{targ} } } }$, respectively. Moreover, 
\begin{align}
	& \hat{H}_{ \text{proj} } \ket{ { \phi }_{i;c_{\rm proj}} } = {E}_{ i , {c}_{ \text{proj} } } \ket{ { \phi }_{i;c_{\rm proj}} } \\
	& \hat{H}_{ \text{targ} } \ket{ {c}_{ \text{targ} } } = {E}_{ {c}_{ \text{targ} } } \ket{ {c}_{ \text{targ} } } \ .
	\label{eq_CC_E_proj}
\end{align}

Matrix elements of the Hamiltonian:
\begin{align}
	{H}_{ c' , c } ( r' , r ) &= \braket{ { \phi }_{ i' }^{ \text{rad} } , c' | \hat{H} | { \phi }_{i}^{ \text{rad} } , c } \nonumber \\
	&= \sum_{ i,i' = 1 }^{N} \frac{ {u}_{ i' } ( r' ) }{ r' } \frac{ {u}_{i} (r) }{r} {H}_{ c' , c } ( i' , i ) 
	\label{eq_CC_ME_Hamilt_Berggren_basis}
\end{align}
and of the norm:
\begin{align}
	{N}_{ c' , c } ( r' , r ) &= \braket{ { \phi }_{ i' }^{ \text{rad} } , c' | { \phi }_{i}^{ \text{rad} } , c } \nonumber \\
	&= \sum_{ i,i' = 1 }^{N} \frac{ {u}_{ i' } ( r' ) }{ r' } \frac{ {u}_{i} (r) }{r} {N}_{ c' , c } ( i' , i ) 
	\label{eq_CC_ME_N_Berggren_basis}
\end{align}
are evaluated using the expansion \eqref{xxx}.

In the calculation of sums in Eqs. \eqref{eq_CC_ME_Hamilt_Berggren_basis} and \eqref{eq_CC_ME_N_Berggren_basis}, four cases have to be considered. In the first case: ${ i < {i}_{ \text{max} } }$ and ${ i' < {i}_{ \text{max} } }$, the matrix elements are calculated in terms of Slater determinants to take into account the antisymmetry.
In the second and third cases: ${ i < {i}_{ \text{max} } }$ and ${ i' \geq {i}_{ \text{max} } }$ and ${ i \geq {i}_{ \text{max} } }$ and ${ i' < {i}_{ \text{max} } }$ which are symmetric with respect to the exchange of ${ i }$ and ${ i' }$, the matrix elements are equal zero because Berggren states ${ \ket{ { \phi }_{i;c_{\rm proj}} } }$ and ${ \ket{ { \phi }_{ i';c_{\rm proj}} } }$ with ${ i \geq {i}_{ \text{max} }}$ or ${ i' \geq {i}_{ \text{max} } }$ are orthogonal to all target states.
In the last case: ${ i \geq {i}_{ \text{max} } }$ and ${ i' \geq {i}_{ \text{max} } }$, there is no antisymmetry and only terms with ${ i = i' }$ are non-zero. One obtains:
\begin{align}
	{H}_{ c' , c } ( r' , r ) &= - \frac{ { \hbar }^{2} }{ 2 \mu } \left( \frac{1}{r} \diffn{ ( r \cdot ) }{r}{2} - \frac{ l ( l + 1 ) }{ {r}^{2} } - {k}_{ {c}_{ \text{targ} } }^{2} \right) \nonumber \\
	&\times \frac{ \delta ( r - r' ) }{ {r}^{2} } { \delta }_{ {c}_{ \text{targ} }' , {c}_{ \text{targ} } } + {V}_{ c' , c } ( r' , r ) 
	\label{eq_CC_EM_H_final}
\end{align}
where
${k}_{ {c}_{ \text{targ} } }^{2} = 2 \mu {E}_{ {c}_{ \text{targ} } } /{ { \hbar }^{2} }$ and
the channel-channel coupling potential ${V}_{ c' , c } ( r' , r )$ is given by:
\begin{equation}
	{V}_{ c' , c } ( r' , r ) = {U}_{ \text{basis} } (r) \frac{ \delta ( r - r' ) }{ {r}^{2} } { \delta }_{ {c}_{ \text{targ} }' , {c}_{ \text{targ} } } + \tilde{V}_{ c' , c } ( r' , r )
	\label{eq_CC_ME_Vccp_pot}
\end{equation}
with
\begin{align}
	&\tilde{V}_{ c' , c } ( r' , r ) = \sum_{i,i' = 1 }^{{i}_{ \text{max} }} \frac{ {u}_{ i' } ( r' ) }{ r' } \frac{ {u}_{i} (r) }{r} {H}_{ c' , c } ( i' , i ) \nonumber \\
	&\quad - \sum_{ i = 1 }^{ {i}_{ \text{max} } - 1 } \frac{ {u}_{ i } ( r' ) }{ r' } \frac{ {u}_{i} (r) }{r} ( {E}_{ i , {c}_{ \text{proj} } } { \delta }_{ {c}_{ \text{targ} }' , {c}_{ \text{targ} } } + {E}_{ {c}_{ \text{targ} } } ) \ .
	\label{eq_CC_EM_Vccp_pot_rest}
\end{align}
In the same way, for ${ {N}_{ c' , c } ( r' , r ) }$ one obtains:
\begin{equation}
	{N}_{ c' , c } ( r' , r ) = \frac{ \delta ( r - r' ) }{ {r}^{2} } { \delta }_{ {c}_{ \text{targ} }' , {c}_{ \text{targ} } } + \tilde{N}_{ c' , c } ( r' , r )
	\label{eq_CC_EM_N_final}
\end{equation}
with:
\begin{align}
	\tilde{N}_{ c' , c } ( r' , r ) &= \sum_{ i,i' = 1 }^{{i}_{ \text{max} }} \frac{ {u}_{ i' } ( r' ) }{ r' } \frac{ {u}_{i} (r) }{r} {N}_{ c' , c } ( i' , i ) \nonumber \\
	&- \sum_{ i = 1 }^{ {i}_{ \text{max} } - 1 } \frac{ {u}_{ i } ( r' ) }{ r' } \frac{ {u}_{i} (r) }{r} { \delta }_{ {c}_{ \text{targ} }' , {c}_{ \text{targ} } } \ .
	\label{eq_CC_EM_Nccp_rest}
\end{align}

\subsection{Orthogonalization of the channel states}
\label{subsubsec_CC_GSM_orthog}

The CC formalism leads to a generalized eigenvalue problem because different channel basis states are non-orthogonal.
The non-orthogonality of channel states comes from the antisymmetry between the projectile and target states.
To formulate GSM-CC equations as the generalized eigenvalue problem, one should express Eq. \eqref{eq_CC_eqs_general} in the orthogonal channel basis ${ \{ \ket{ r , c }_{o} \} }$:
\begin{equation}
	{}_{o}\braket{ r' , c' | r , c }_{o} = \frac{ \delta ( r' - r ) }{ {r}^{2} } { \delta }_{ c' c } \ .
	\label{eq_CC_ortho_channel_basis_braket}
\end{equation}
The transformation from the non-orthogonal channel basis ${ \{ \ket{ r , c }\}}$ to the orthogonal one ${ \{ \ket{ r , c }_{o}\} }$ is given by the overlap operator ${ \hat{O} }$ such that: $\ket{ r , c } = \hat{O}^{ \frac{1}{2} } \ket{ r , c }_{o} \label{eq_CC_non_ortho_to_ortho_channel_states}$. The CC equations \eqref{eq_CC_eqs_general} written in the orthogonal basis are:
\begin{align}
	\sumint\limits_{c} \intt_{0}^{ \infty } dr \, {r}^{2} &( {}_{o}\braket{ r' , c' | \hat{H}_{o} | r , c }_{o} - E {}_{o}\braket{ r' , c' | \hat{O} | r , c }_{o} ) \nonumber \\
	&\times {}_{o}\braket{ r , c | { \Psi }_{o} } = 0 \ ,
	\label{eq_CC_eqs_general_clear_ortho}
\end{align}
where
${}_{o}\braket{ r' , c' | \hat{H}_{o} | r , c }_{o} = \braket{ r' , c' | \hat{H} | r , c } $,~
${}_{o}\braket{ r' , c' | \hat{O} | r , c }_{o} = \braket{ r' , c' | r , c }$, and 
$ {}_{o}\braket{ r , c | { \Psi }_{o} } = \braket{ r , c | \Psi } 
\label{eq_CC_u_c_non_ortho_to_ortho}$.
The transformation of this generalized eigenvalue problem into a standard eigenvalue problem is achieved with a substitution: 
${ \ket{ \Phi } = \hat{O} \ket{ \Psi } }$. One obtains:
\begin{equation}
	\sumint\limits_{c} \intt_{0}^{ \infty } dr \, {r}^{2} ( {}_{o}\braket{ r' , c' | \hat{H} | r , c }_{o} - E {}_{o}\braket{ r' , c' | r , c }_{o} ) {}_{o}\braket{ r , c | \Phi } = 0
	\label{eq_CC_eqs_general_clear_ortho_again}
\end{equation}
with ${}_{o}\braket{ r , c | \Phi } = \braket{ r , c | \hat{O}^{ \frac{1}{2} } | \Psi } \equiv {w}_{c} (r) r \label{eq_CC_w_c_def}$.
In the non-orthogonal channel basis, these CC equations become:
\begin{equation}
	\sumint\limits_{c} \intt_{0}^{ \infty } dr \, {r}^{2} \braket{ r' , c' | \hat{H}_{m} | r , c } \frac{ {w}_{c} (r) }{r} = E \frac{ {w}_{ c' } ( r' ) }{ r' } \ ,
	\label{eq_CC_final}
\end{equation}
with ${}_{o}\braket{ r' , c' | \hat{H} | r , c }_{o} \equiv \braket{ r' , c' | \hat{H}_{m} | r , c }
\label{eq_CC_Hamilt_modif_def}$, where ${ \hat{H}_{m} = \hat{O}^{ - \frac{1}{2} } \hat{H} \hat{O}^{ - \frac{1}{2} } }$ is the modified Hamiltonian. 

Matrix elements of ${ \hat{H}_{m} }$ are calculated using the expansion \eqref{xxx} as described in Sec. \ref{subsubsec_CC_GSM_matrix_elts}.
In order to have a more precise treatment of the antisymmetry in the calculation of matrix elements of ${ \hat{H}_{m} }$, we introduce a new operator ${ \hat{ \Delta } }$:
$\hat{O}^{ - \frac{1}{2} } = \hat{ \Delta } + \hat{\bf 1}
\label{eq_CC_Delta_def}$,
which is associated with the part of ${ \hat{O}^{ - \frac{1}{2} } }$ acting on the low-energy channel states.
Then, instead of calculating the matrix elements of ${ \hat{H}_{m} }$ directly, it is possible to calculate them as:
\begin{equation}
	{H}_{m} = ( \Delta + \hat{\bf 1} ) H ( \Delta + \hat{\bf 1} ) = H + H \Delta + \Delta H + \Delta H \Delta
	\label{eq_CC_H_m_trick}
\end{equation}
In this formulation, the non-antisymmetrized terms are taken into account exactly with the identity operator.
Inserting \eqref{eq_CC_H_m_trick} in CC equations \eqref{eq_CC_final} and replacing matrix elements ${ \braket{ r' , c' | \hat{H} | r , c } }$ using \eqref{eq_CC_EM_H_final} and \eqref{eq_CC_ME_Vccp_pot}, one obtains the CC equations for the reduced radial wave functions ${ {w}_{c} (r) / r }$:
\begin{widetext}
	\begin{align}
		& \left( - \frac{ { \hbar }^{2} }{ 2 \mu } \left( \frac{1}{r} \diffn{ ( r \cdot ) }{r}{2} - \frac{ l ( l + 1 ) }{ {r}^{2} } \right) + {V}_{c}^{ ( \text{loc} ) } (r) \right) \frac{ {w}_{c} (r) }{r} \frac{ \delta ( r - r' ) }{ {r}^{2} } { \delta }_{ {c}_{ \text{targ} }' , {c}_{ \text{targ} } } + \sum_{ c' } \intt_{0}^{ \infty } dr' \, r { r' }^{2} \frac{ {V}_{ c , c' }^{ ( \text{non-loc} ) } ( r , r' ) }{ r r' } \frac{ {w}_{ c' } ( r' ) }{ r' } \nonumber \\
		&= ( E - {E}_{ {c}_{ \text{targ} } } ) \frac{ \delta ( r - r' ) }{ {r}^{2} } \frac{ {w}_{c} (r) }{r} { \delta }_{ {c}_{ \text{targ} }' , {c}_{ \text{targ} } }
		\label{eq_CC_modif_Eqs_full_2}
	\end{align}
\end{widetext}
with the local potential ${ {V}_{c}^{ ( \text{loc} ) } (r) = {U}_{ \text{basis} } (r) }$ which may depend on the channel ${ c }$, and the non-local potential:
\begin{align}
	\frac{1}{ r' r } {V}_{ c' , c }^{ ( \text{non-loc} ) } &( r' , r ) = \tilde{V}_{ c' , c } ( r' , r ) + \braket{ r' , c' | \hat{H} \hat{ \Delta } | r , c } \nonumber \\
	&+ \braket{ r' , c' | \hat{ \Delta } \hat{H} | r , c } + \braket{ r' , c' | \hat{ \Delta } \hat{H} \hat{ \Delta } | r , c } \ .
	\label{eq_CC_non_local_pot_final}
\end{align}
The radial channel wave functions ${ {u}_{c} (r) / r }$ are then obtained from the solutions of Eq. \eqref{eq_CC_modif_Eqs_full_2} using the equation:
\begin{equation}
	\frac{ {u}_{c} (r) }{r} = \frac{ {w}_{c} (r) }{r} + \sum_{ c' } \intt_{0}^{ \infty } dr' \, { r' }^{2} \braket{ r , c | \hat{O}^{ \frac{1}{2} } \hat{ \Delta } \hat{O}^{ \frac{1}{2} } | r' , c' } \frac{ {w}_{ c' } ( r' ) }{ r' } \ .
	\label{eq_uc_wc_link}
\end{equation}

\subsection{Solution of the GSM-CC equations}
\label{subsubsec_CC_GSM_equiv_pot_meth}

CC equations \eqref{eq_CC_modif_Eqs_full_2} contain a non-local potential which has to be treated using a generalization of the method of the equivalent potential \cite{michel09_147,jaganathen14_988}. The basic idea is to find the equivalent local potential ${V}_{ c , c' }^{ ( \text{eq} ) } (r)$ and the source term ${S}_{c} (r)$ which would replace local ${V}_{c}^{ ( \text{loc})}(r)$ and non-local ${V}_{ c , c' }^{ ( \text{non-loc} ) } ( r , r' )$ potentials in \eqref{eq_CC_modif_Eqs_full_2}.
Such an equivalent potential is defined by:
\begin{align}
	&{V}_{ c , c' }^{ ( \text{eq} ) } (r) = {V}_{c}^{ ( \text{loc} ) } (r) { \delta }_{ c' , c } \nonumber \\
	&\quad + \frac{ 1 - {F}_{ c' } (r) }{ {w}_{ c' } (r) } \sum_{ c' } \intt_{0}^{ \infty } dr' \, {V}_{ c , c' }^{ ( \text{non-loc} ) } ( r , r' ) {w}_{ c' } ( r' )
	\label{eq_CC_equi_pot_def}
\end{align}
and a corresponding source term is:
\begin{equation}
	{S}_{c} (r) = {F}_{ c' } (r) \sum_{ c' } \intt_{0}^{ \infty } dr' \, {V}_{ c , c' }^{ ( \text{non-loc} ) } ( r , r' ) {w}_{ c' } ( r' ) \ .
	\label{eq_CC_source_def}
\end{equation}
${ {F}_{c} (r) }$ in Eqs. \eqref{eq_CC_equi_pot_def}, \eqref{eq_CC_source_def} is the smoothing function:
\begin{equation}
	{F}_{c} (r) = \exp - \alpha { \left| \frac{ {w}_{c} (r) }{ {w}_{c}' (r) } \right| }^{2}  \left( 1 - \exp - \alpha { \left| \frac{ {w}_{c}^{ \text{asymp} } (r) }{ {w}_{c} (r) } - 1 \right| }^{2}  \right)
	\label{eq_CC_equiv_pot_smooth_func_def}
\end{equation}
to cancel divergences of the equivalent potential ${ {V}_{ c , c' }^{ ( \text{eq} ) } (r) }$ close to the zeroes of ${ {w}_{c} (r) }$.
In this expression:
${w}_{c}' (r) = {w}_{c} (r)/{r}
\label{eq_CC_derivative_wc}$,
and ${ {w}_{c}^{ \text{asymp} } }$ is the asymptotic form of ${ {w}_{c} (r) }$ when ${ r \sim 0 }$. Typically, the value of ${ \alpha }$ varies in the interval $10 < \alpha < 100$. 

With these substitutions, the GSM-CC equations \eqref{eq_CC_modif_Eqs_full_2} become:
\begin{align}
	\diffn{ {w}_{c} (r) }{r}{2} &= \left( \frac{ l ( l + 1 ) }{ {r}^{2} } - {k}_{c}^{2} \right) {w}_{c} (r) \nonumber \\
	&+ \frac{ 2 \mu }{ { \hbar }^{2} } \left( \sum_{ c' } {V}_{ c , c' }^{ ( \text{eq , sy} ) } (r) {w}_{ c' } (r) + {S}_{c}^{ ( \text{sy} ) } (r) \right)
	\label{eq_CC_modif_Eqs_full_6}
\end{align}
where:
${k}_{c}^{2} = { 2 \mu } ( E - {E}_{ {c}_{ \text{targ} } } )/{ { \hbar }^{2} }
\label{eq_CC_kc_def}$.
Eqs. \eqref{eq_CC_modif_Eqs_full_6} are solved iteratively to determine the equivalent potential, the source term, and the mutally orthogonal radial wave functions ${ {w}_{c} (r) }$. Starting point for solving these equations is provided by a set of radial channel wave functions ${ \{ {w}_{c} (r) \} }$ obtained by the diagonalization of GSM-CC equations \eqref{eq_CC_final} in the Bergen basis of channels. Diagonalization of CC equations in the Bergen basis was also considered in Ref. \cite{betan14_825}.
Note that it is numerically more convenient to express the potential ${V}_{ c , c' }^{ ( \text{non-loc} ) } ( r , r' )$ of Eq.(\eqref{eq_CC_modif_Eqs_full_2}) in a basis of harmonic oscillator states,
as ${V}_{ c , c' }^{ ( \text{non-loc} ) } ( r , r' )$ is short-range. For this, it is sufficient to replace all occurrences of Berggren basis functions $u_i(r)/r$ by harmonic oscillator states overlaps $\braket{u_i|u_n^{(HO)}}$
in Eqs.(\ref{eq_CC_EM_Vccp_pot_rest},\ref{eq_CC_EM_Nccp_rest}), where $\ket{u_n^{(HO)}}$ is a harmonic oscillator state.

\section{The radiative capture process}
\label{sec_rad_cap_cross_sec}

We now discuss the calculation of proton/neutron radiative capture cross sections using the antisymmetrized initial and final GSM wave functions. 
The differential cross section for a proton or neutron radiative capture can be calculated from the Fermi golden rule, which relates the cross section to the matrix elements of a transition operator between an initial state ${ \ket{i} }$ of energy ${ {E}_{i} }$ and a final state ${ \ket{f} }$ of energy ${ {E}_{f} }$. The differential cross section is given by:
% \cite{dubovichenko97_316}:
\begin{widetext}
	\begin{align}
		\ddiff{ \sigma }{ { \Omega }_{ \gamma } } &= \frac{1}{ 8 \pi } \left( \frac{ {k}_{ \gamma } }{k} \right) \left( \frac{ {e}^{2} }{ \hbar c } \right) \left( \frac{ { \mu }_{u} {c}^{2} }{ \hbar c } \right) \frac{1}{ 2 s + 1 } \frac{1}{2 {J}_{ \text{targ} } + 1} \nonumber \\
		&\times \sum_{ \substack{ {M}_{i} , {M}_{f} , \\ {M}_{ \text{targ} } , {M}_{L} , \\ P , {m}_{s} } } { \left| \sum_{L} {i}^{L} \sqrt{ 2 \pi ( 2 L + 1 ) } \left( \frac{ {k}_{ \gamma }^{L} }{k} \right) \sqrt{ \frac{ L + 1 }{L} } \frac{P}{ ( 2 L + 1 ) !! } {D}^{L}_{ {M}_{L} P} ( { \varphi }_{ \gamma } , { \theta }_{ \gamma } , 0 ) \braket{ { \Psi }_{f} ( {J}_{f} , {M}_{f} ) | \hat{ \mathcal{M} }_{ L , {M}_{L} } | { \Phi }_{i} ( {M}_{i} ) } \right| }^{2} \nonumber \\
		&= \frac{1}{ 2 \pi } \left( \frac{ {k}_{ \gamma } }{k} \right) \left( \frac{ {e}^{2} }{ \hbar c } \right) \left( \frac{ { \mu }_{u} {c}^{2} }{ \hbar c } \right) \frac{1}{ 2 s + 1 } \frac{1}{2 {J}_{ \text{targ} } + 1} \sum_{ \substack{ {M}_{i} , {M}_{f} , \\ {M}_{ \text{targ} } , {M}_{L} , \\ P , {m}_{s} } } { \left| \sum_{L} {g}^{L}_{ {M}_{L} , P } ( k , {k}_{ \gamma } , { \varphi }_{ \gamma } , { \theta }_{ \gamma } ) \braket{ { \Psi }_{f} ( {J}_{f} , {M}_{f} ) | \hat{ \mathcal{M} }_{ L , {M}_{L} } | { \Phi }_{i} ( {M}_{i} ) } \right| }^{2} 
		\label{eq_rad_cap_diff_cross_sec_0}
	\end{align}
\end{widetext}
where:
\begin{align}
	{g}^{L}_{ {M}_{L} , P } &(k , {k}_{ \gamma } , { \varphi }_{ \gamma } , { \theta }_{ \gamma } ) = {i}^{L} \sqrt{ 2 \pi ( 2 L + 1 ) } \left( \frac{ {k}_{ \gamma }^{L} }{k} \right) \nonumber \\
	&\times \sqrt{ \frac{ L + 1 }{L} } \frac{P}{ ( 2 L + 1 ) !! } {D}^{L}_{ {M}_{L} P} ( { \varphi }_{ \gamma } , { \theta }_{ \gamma } , 0)  \ .
	\label{eq_rad_cap_ang_factor}
\end{align}
In the above expressions, ${ {k}_{ \gamma } }$ (in units of ${ { \text{fm} }^{ -1 } }$) is the linear momentum of the emitted photon: 
${ {k}_{ \gamma } = ( {E}_{f} - {E}_{i} ) / ( \hbar c ) }$, ${ {e}^{2} / ( \hbar c ) }$ is the electromagnetic coupling constant,
${ k }$ (in units of ${ { \text{fm} }^{ -1 } }$) is the linear momentum of the incoming proton in the CM reference frame,
${ { \mu }_{u} {c}^{2} }$ (in MeV) is the reduced mass of the total system of ${ A }$ nucleons, ${ s }$ is the spin of the proton, ${ {J}_{ \text{targ} } }$ is the total angular momentum of the target, ${ P = \pm 1 }$ is the polarization of the photon, ${ L }$ and ${ {M}_{L} }$ are the multipoles and multipole projections of the photon. Moreover,  ${ {D}^{L}_{ {M}_{L} P }( { \varphi }_{ \gamma } , { \theta }_{ \gamma } , 0 ) }$ is the Wigner D-matrix depending on the angular variables ${ { \theta }_{ \gamma } }$ and ${ { \varphi }_{ \gamma } }$ of the photon, and  ${ \hat{ \mathcal{M} }_{ L , {M}_{L} } }$ is the electromagnetic transition operator.
The final state ${ \ket{f} }$ corresponds to the GSM-CC state ${ \ket{ { \Psi }_{f} ( {J}_{f} , {M}_{f} ) } }$ of a total angular momentum ${ {J}_{f} }$ and a projection ${ {M}_{f} }$.
The initial state ${ \ket{i} }$ has a fixed value of the total angular momentum projection ${ {M}_{i} }$ and is denoted ${ \ket{ { \Phi }_{i} ( {M}_{i} ) } }$:
\begin{align}
	\ket{ { \Phi }_{i} ( {M}_{i} ) } &= \sum_{ {J}_{i} , {c}_{e} } {i}^{ {l}_{ {c}_{e} } } {e}^{ i { \sigma }_{ {l}_{ {c}_{e} } } } \sqrt{ 2 {l}_{ {c}_{e} } + 1 } \ket{ { \Psi }_{i} ( {J}_{i} , {M}_{i} , {c}_{e} ) } \nonumber \\
	&\times \braket{ {l}_{ {c}_{e} } , 0 , s , {m}_{s} | ( {l}_{ {c}_{e} } , s ) {j}_{ {c}_{e} } , {m}_{s} } \nonumber \\
	&\times \braket{ {j}_{ {c}_{e} } , {m}_{s} , {J}_{ \text{targ} } , {M}_{ \text{targ} } | ( {j}_{ {c}_{e} } , {J}_{ \text{targ} } ) {J}_{i} , {M}_{i} }
	\label{eq_rad_cap_initial_state}
\end{align}
where ${ \ket{ { \Psi }_{i} ( {J}_{i} , {M}_{i} , {c}_{e} ) } }$ is the initial GSM-CC state with a total angular momentum ${ {J}_{i} }$ and an entrance channel quantum numbers ${ {c}_{e} }$. Each set of quantum numbers ${ {c}_{e} }$ corresponds to a different channel 
${ c }$.
This state can be expressed in the channel basis as:
$$\ket{ { \Psi }_{i} ( {J}_{i} , {M}_{i} , {c}_{e} ) } = \sum_{c} \ket{ { \Psi }_{i} ( {J}_{i} , {M}_{i} , {c}_{e} ) }_{c}
\label{eq_rad_cap_CC_initial_state}$$.
\begin{widetext}
	Thus, the differential cross section (in units of ${ \text{fm}^{2} }$) writes:
	\begin{align}
	  \ddiff{ \sigma }{ { \Omega }_{ \gamma } } &= \frac{1}{2 \pi} \left( \frac{ {k}_{\gamma} }{k} \right) \left( \frac{ {e}^{2} }{\hbar c} \right) \left( \frac{ {\mu}_{u} {c}^{2} }{\hbar c} \right) \frac{1}{2s+1} \frac{1}{2 {J}_{\text{targ}} + 1} \nonumber \\
		&\times \sum_{ \substack{ {M}_{i} , {M}_{f} , \\ P , {m}_{s} , \\ {M}_{\text{targ}} , {M}_{L} } } { \left| \sum_{L} {g}^{L}_{ {M}_{L} , P } (k , {k}_{ \gamma } , { \varphi }_{ \gamma } , { \theta }_{ \gamma } ) \sum_{ {J}_{i} , {c}_{e} } \braket{ {J}_{f} {M}_{f} | \mathcal{M}_{L , {M}_{L} } | { ( {J}_{i} {M}_{i} ) }_{ {c}_{e} } } \braket{ {l}_{ {c}_{e} } 0 s {m}_{s} | {j}_{ {c}_{e} } {m}_{s} } \braket{ {j}_{ {c}_{e} } {m}_{s} {J}_{\text{targ}} {M}_{\text{targ}} | {J}_{i} {M}_{i} } \right| }^{2} \nonumber \\
		&= \frac{1}{2 \pi} \left( \frac{ {k}_{\gamma} }{k} \right) \left( \frac{ {e}^{2} }{\hbar c} \right) \left( \frac{ {\mu}_{u} {c}^{2} }{\hbar c} \right) \frac{1}{2s+1} \frac{1}{2 {J}_{\text{targ}} + 1} \times \sum_{ \substack{ {M}_{i} , {M}_{f} , \\ P , {m}_{s} , \\ {M}_{\text{targ}} , {M}_{L} } } \sum_{ \substack{ L , L' , \\ {J}_{i} , {J}_{i}' , \\ {c}_{e} , {c}_{e}' } } \left( \frac{}{} {g}^{L}_{ {M}_{L} , P } (k , {k}_{ \gamma } , { \varphi }_{ \gamma } , { \theta }_{ \gamma } ) {g}^{L'}_{ {M}_{L} , P } (k , {k}_{ \gamma } , { \varphi }_{ \gamma } , { \theta }_{ \gamma } ) \right. \nonumber \\
		&\times \braket{ {J}_{f} {M}_{f} | \mathcal{M}_{L , {M}_{L} } | { ( {J}_{i} {M}_{i} ) }_{ {c}_{e} } } \braket{ {J}_{f} {M}_{f} | \mathcal{M}_{L' , {M}_{L} } | { ( {J}_{i}' {M}_{i} ) }_{ {c}_{e}' } } \times \braket{ {l}_{ {c}_{e} } 0 s {m}_{s} | {j}_{ {c}_{e} } {m}_{s} } \braket{ {l}_{ {c}_{e}' } 0 s {m}_{s} | {j}_{ {c}_{e}' } {m}_{s} } \nonumber \\
		&\times \left. \braket{ {j}_{ {c}_{e} } {m}_{s} {J}_{\text{targ}} {M}_{\text{targ}} | {J}_{i} {M}_{i} } \braket{ {j}_{ {c}_{e}' } {m}_{s} {J}_{\text{targ}} {M}_{\text{targ}} | {J}_{i}' {M}_{i} } \frac{}{} \right) \ .
		\label{eq_rad_cap_diff_cross_sec_2}
	\end{align}
\end{widetext}
The operator ${ \hat{ \mathcal{M} }_{ L , {M}_{L} } }$ separates into an electric part ${ \hat{ \mathcal{M} }_{ L , {M}_{L} }^{E} }$ and a magnetic part ${ \hat{ \mathcal{M} }_{ L , {M}_{L} }^{M} }$. Formulae for the operators ${ \hat{ \mathcal{M} }_{ L , {M}_{L} }^{E} }$ and ${ \hat{ \mathcal{M} }_{ L , {M}_{L} }^{M} }$ are given in Appendix \ref{annexe_rad_cap_EM_operators}.

\subsection{Calculation of many-body matrix elements of the electromagnetic operators}
\label{sec_rad_cap_method}

The main difficulty in the calculation of matrix elements comes from the infinite-range of the electromagnetic operators and the antisymmetry of the GSM-CC states. Indeed, direct calculation of these matrix elements in the Berggren basis is not possible because they diverge even using the exterior complex scaling method. 

If one neglects antisymmetry in the channel state ${ \ket{ r , c } }$:	
\begin{equation}
	\ket{ r , c } = \ket{r} \otimes \ket{c} = \ket{r} \otimes { [ \ket{ {J}_{ \text{targ} , c } , {M}_{ \text{targ} , c } } \otimes \ket{ {l}_{c} , {s}_{c} ; {j}_{c} , {m}_{ {j}_{c} } } ] }^{J}_{M} 
	\label{eq_rad_cap_channel_state_nas}
\end{equation}
then the overlap between a bound state or a narrow resonance and a scattering state converges using the exterior complex-scaling method. In the above expression, ${ {J}_{ \text{targ} , c } }$ is the angular momentum of the target in a channel ${ c }$ with a projection ${ {M}_{ \text{targ} , c } }$, ${ {l}_{c} }$ is the orbital momentum of the projectile, ${ {s}_{c} }$ its spin and ${ {j}_{c} }$ its total angular momentum with a projection ${ {m}_{ {j}_{c} } }$. The antisymmetry between the target and the projectile can be neglected only at large distances because the probability that the one-body state of the projectile is occupied by the target nucleon is the lower the smaller is the target density. In this case, the action of a given operator ${ \hat{O}^{L}_{ {M}_{L} }}$ can be defined by considering target nucleons as distinguishable from the projectile nucleons:
\begin{equation}
	\hat{O}^{L}_{ {M}_{L} } = \sum_{ i \in A } \hat{O}^{L}_{ {M}_{L} } ( {r}_{i} , { \Omega }_{i} ) + \hat{O}^{L}_{ {M}_{L} } ( {r}_{ \text{proj} } , { \Omega }_{ \text{proj} } )
	\label{eq_rad_cap_O_nas}
\end{equation}
The first sum acts only on target nucleons whereas the second term acts on a projectile. Obviously, this approximation is not valid for a target in the continuum state.

The calculation of matrix elements of the electromagnetic operators goes as follows.
The matrix elements are expressed as the sum of a non-antisymmetrized (${ \text{nas} }$) part and its complement:
\begin{eqnarray}
	\braket{ { \Psi }_{f} || \hat{O}^{L} || { \Psi }_{i} } &=& \braket{ { \Psi }_{f} || \hat{O}^{L} || { \Psi }_{i} }_{ \text{nas} } \nonumber \\
	&+& \left( \braket{ { \Psi }_{f} || \hat{O}^{L} || { \Psi }_{i} } - \braket{ { \Psi }_{f} || \hat{O}^{L} || { \Psi }_{i} }_{ \text{nas} } \right)
	\label{eq_rad_cap_method_nas_plus_rest}
\end{eqnarray}
The calculation of this complement is achieved by separating the operator ${ \hat{O}^{L} }$ into a short-range part ${ \hat{O}_{<}^{L} }$ and a long-range part ${ \hat{O}_{>}^{L} }$. Then the symmetrized and antisymmetrized matrix elements write:
\begin{eqnarray}
	&& \braket{ { \Psi }_{f} || \hat{O}^{L} || { \Psi }_{i} } = \braket{ { \Psi }_{f} || \hat{O}_{<}^{L} || { \Psi }_{i} } + \braket{ { \Psi }_{f} || \hat{O}_{>}^{L} || { \Psi }_{i} } \\
	&& \braket{ { \Psi }_{f} || \hat{O}^{L} || { \Psi }_{i} }_{ \text{nas} } = \braket{ { \Psi }_{f} || \hat{O}_{<}^{L} || { \Psi }_{i} }_{ \text{nas} } + \braket{ { \Psi }_{f} || \hat{O}_{>}^{L} || { \Psi }_{i} }_{ \text{nas} } \nonumber \\
	\label{eq_rad_cap_method_rest_calc}
\end{eqnarray}
At large distances, the antisymmetry is not crucial and thus the matrix element ${ \braket{ { \Psi }_{f} || \hat{O}_{>}^{L} || { \Psi }_{i} } }$ can be approximated by ${ \braket{ { \Psi }_{f} || \hat{O}_{>}^{L} || { \Psi }_{i} }_{ \text{nas} } }$.
The remaining term is basically a short-range part which can be expanded in the HO basis. One obtains:
\begin{eqnarray}
	\braket{ { \Psi }_{f} || \hat{O}^{L} || { \Psi }_{i} } &=& \braket{ { \Psi }_{f} || \hat{O}^{L} || { \Psi }_{i} }_{ \text{nas} } \nonumber \\
	&+& \braket{ { \Psi }_{f} || \hat{O}_{<}^{L} || { \Psi }_{i} }^{ \text{HO} } - \braket{ { \Psi }_{f} || \hat{O}_{<}^{L} || { \Psi }_{i} }_{ \text{nas} }^{ \text{HO} } \nonumber \\
	\label{eq_rad_cap_method}
\end{eqnarray}

The matrix element ${ \braket{ { \Psi }_{f} || \hat{O}^{L} || { \Psi }_{i} }_{ \text{nas} } }$ is not antisymmetrized. We may write the operator ${ \hat{O}^{L} }$ (Eq. \eqref{eq_rad_cap_O_nas}) as: ${ \hat{O}_{ \text{targ} }^{L} + \hat{O}_{ \text{proj} }^{L} }$, where ${ \hat{O}_{ \text{targ} }^{L} }$ acts only on the target state and ${ \hat{O}_{ \text{proj} }^{L} }$ on the projectile state. In this case, matrix elements of the electromagnetic operator acting on target states are:
\begin{widetext}
	\begin{align}
		{}_{ {c}_{f} } \braket{ { \Psi }_{f} || \hat{O}_{ \text{targ} }^{L} || { \Psi }_{i} }_{ {c}_{i} } &= \intt_{0}^{ \infty } dr \, {r}^{2} \frac{ {u}_{ {c}_{f} } (r) }{r} \intt_{0}^{ \infty } dr' \, { r' }^{2} \frac{ {u}_{ {c}_{i} } ( r' ) }{ r' } \braket{ r | r' } \nonumber \\
		&\times \braket{ {l}_{ {c}_{f} } , {s}_{ {c}_{f} } ; {j}_{ {c}_{f} } , {m}_{ {j}_{ {c}_{f} } } | {l}_{ {c}_{i} } , {s}_{ {c}_{i} } ; {j}_{ {c}_{i} } , {m}_{ {j}_{ {c}_{i} } } } \braket{ {J}_{ {T}_{ {c}_{f} } } || \hat{O}_{ \text{targ} }^{L} || {J}_{ {T}_{ {c}_{i} } } } \nonumber \\
		&= { ( -1 ) }^{ {J}_{ {T}_{f} } + {j}_{f} + {J}_{i} + L} \sqrt{ (2 {J}_{f} + 1) (2 {J}_{i} + 1) } \WignerSIXj{ {J}_{ {T}_{f} } }{ {J}_{ {T}_{i} } }{L}{ {J}_{i} }{ {J}_{f} }{ {j}_{i} } \braket{ {J}_{ {T}_{f} } || \hat{O}^{L} || {J}_{ {T}_{i} } } \nonumber \\
		&\times { \delta }_{ {l}_{i} {l}_{f} } { \delta }_{ {j}_{i} {j}_{f} } \intt_{0}^{ \infty } dr \, {u}_{ {c}_{f} } (r) {u}_{ {c}_{i} } (r)
		\label{eq_rad_cap_nas_ME_target}
	\end{align}
\end{widetext}
where ${ \braket{ r | r' } = { \delta }_{ r , r' } / {r}^{2} }$, and ${ {c}_{i} }$ and ${ {c}_{f} }$ denote
initial and final channels, respectively. No exterior complex scaling is necessary to calculate the radial overlap in the above expression because ${ {u}_{ {c}_{i} }(r) }$ is the scattering wave function of a real energy and ${ {u}_{ {c}_{f} }(r) }$ is the bound state wave function.
Similarly, matrix elements of the electromagnetic operator acting on the projectile states are:
\begin{widetext}
	\begin{align}
		{}_{ {c}_{f} } \braket{ { \Psi }_{f} || \hat{O}_{ \text{proj} }^{L} || { \Psi }_{i} }_{ {c}_{i} } &= \intt_{0}^{ \infty } dr \, {r}^{2} \frac{ {u}_{ {c}_{f} } (r) }{r} \intt_{0}^{ \infty } dr' \, { r' }^{2} \frac{ {u}_{ {c}_{i} } ( r' ) }{ r' } \braket{ r | r' } \nonumber \\
		&\times \braket{ {J}_{ {T}_{ {c}_{f} } } , {M}_{ {T}_{ {c}_{f} } } | {J}_{ {T}_{ {c}_{i} } } , {M}_{ {T}_{ {c}_{i} } } } \braket{ ( {l}_{ {c}_{f} } , {s}_{ {c}_{f}  } ) {j}_{ {c}_{f} } || \hat{O}_{ \text{proj} }^{L} || ( {l}_{ {c}_{i} } , {s}_{ {c}_{i} } ) {j}_{ {c}_{i} } } \nonumber \\
		&= { \delta }_{ {T}_{i} {T}_{f} } { (-1) }^{ {J}_{ {T}_{i} } + {j}_{i} + {J}_{f} + L} \sqrt{ (2 {J}_{f} + 1) (2 {J}_{i} + 1) } \WignerSIXj{ {j}_{f} }{ {j}_{i} }{L}{ {J}_{i} }{ {J}_{f} }{ {J}_{ {T}_{i} } } \nonumber \\
		&\times \braket{ {u}_{ {c}_{f} } , ( {l}_{ {c}_{f} } , s ) {j}_{ {c}_{f} } || \hat{O}^{L} || {u}_{ {c}_{i} } , ( {l}_{ {c}_{i} } , s ) {j}_{ {c}_{i} } }
		\label{eq_rad_cap_nas_ME_proton}
	\end{align}
\end{widetext}

The antisymmetrized matrix elements ${ \braket{ { \Psi }_{f} || \hat{O}^{L} || { \Psi }_{i} }^{ \text{HO} } }$ in Eq. \eqref{eq_rad_cap_method} are obtained by expressing Berggren basis states in the HO basis. In this case, the reduced radial wave functions ${ {u}_{c} (r) }$ can be written as:
\begin{align}
	\frac{ {u}_{c} (r) }{r} &= \braket{ r | {u}_{c} } \nonumber \\
	&\to \sum_{ n } \braket{ r | {u}_{ n } } \braket{ {u}_{ n } | {u}_{c} } \nonumber \\
	&= \sum_{ n } {u}_{ n } (r) \braket{ {u}_{ n } | {u}_{c} } = \braket{ r | {u}_{c}^{ \text{HO} } } = \frac{ {u}_{c}^{ \text{HO} } (r) }{r} 
	\label{eq_rad_cap_uc_HO}
\end{align}
where ${ \ket{ {u}_{ n } } }$ is the radial HO state and
the channel state ${ \ket{ r , c } }$ can be expressed as:
\begin{align}
	\ket{ r , c } &= \hat{ \mathcal{A} } \left( \ket{r} \otimes \ket{c} \right) \nonumber \\
	&= \hat{ \mathcal{A} } \left( \left( \sum_{n} \braket{ {u}_{n} | r } \ket{ {u}_{n} } \right) \otimes \ket{c} \right) \nonumber \\
	&= \sum_{n} \braket{ {u}_{n} | r } \ket{ {u}_{n} , c } = \frac{ {u}_{c}^{ \text{HO} } (r) }{r} \ket{ {u}_{n} , c } 
	\label{eq_rad_cap_channel_state_Berggren_expansion}
\end{align}
with
$\ket{ {u}_{n} , c } = { \left[ \hat{a}_{ n , {j}_{c} , {m}_{ {j}_{c} } }^{ \dag } \ket{ {J}_{ \text{targ} , c } , {M}_{ \text{targ} , c } } \right] }_{M}^{J}
\label{eq_rad_cap_creation_op_Berggren_basis}$.
Hence, the CC representation of initial and final states in HO basis is:
\begin{align}
	\ket{ \Psi }^{ \text{HO} } = \sumint\limits_{c} \sum_{n} \braket{ {u}_{n} | {u}_{c}^{ \text{HO} } } \ket{ {u}_{n} , c }
	\label{eq_rad_cap_CC_state_fixed_c_HO_final}
\end{align}
and the antisymmetrized matrix elements of the electromagnetic operator are:
\begin{align}
	&{}^{ \text{HO} } \braket{ { \Psi }_{f} || \hat{O}^{L} || { \Psi }_{i} }^{ \text{HO} } = \sum_{ {c}_{i} , {c}_{f} } \sum_{ {n}_{i} , {n}_{f} } \braket{ {u}_{ {c}_{i} }^{ \text{HO} } | {u}_{ {n}_{i} } } \braket{ {u}_{ {n}_{f} } | {u}_{ {c}_{f} }^{ \text{HO} } } \nonumber \\
	&\quad \times { \left[ \bra{ {J}_{ {T}_{ {c}_{f} } } } \hat{a}_{ {n}_{f} , {j}_{ {c}_{f} } } \right] }_{M}^{J} \hat{O}_{ {M}_{L} }^{L} { \left[ \hat{a}_{ {n}_{i} , {j}_{ {c}_{i} } }^{ \dag } \ket{ {J}_{ {T}_{ {c}_{i} } } } \right] }_{M}^{J}
	\label{eq_rad_cap_ME_HO_final}
\end{align}
HO expansion is hereby justified by the fact that the target states are localized.

The last many-body matrix element in Eq.\eqref{eq_rad_cap_method}: ${ {}^{ \text{HO} } \braket{ { \Psi }_{f} || \hat{O}^{L} || { \Psi }_{i} }_{ \text{nas} }^{ \text{HO} } }$, is calculated using Eqs. \eqref{eq_rad_cap_nas_ME_target} and \eqref{eq_rad_cap_nas_ME_proton} and replacing ${ {u}_{c} (r) }$ by ${ {u}_{c}^{ \text{HO} } (r) }$ (see Eq. \eqref{eq_rad_cap_uc_HO}).

\section{Results of GSM-CC calculations for the $^7$Be(p,$\gamma$)$^8$B reaction} 
\label{sec3}

GSM-CC calculations are done in COSM coordinates but the radiative capture cross section is expressed in the CM reference frame. The initial energy is:
${E}_{i}^{ \text{(COSM)} } = {E}_{ \text{proj} }^{ \text{(COSM)} } + {E}_{T}^{ \text{(COSM)} }
\label{eq_rad_cap_tot_initial_energy}$,
where ${ {E}_{i}^{ \text{(COSM)} } }$, ${ {E}_{ \text{proj} }^{ \text{(COSM)} } }$ and ${ {E}_{T}^{ \text{(COSM)} } }$ are the total energy, the projectile energy, and the GSM target binding energy, respectively. All energies are calculated in the COSM coordinate system.
The link between the projectile energies in COSM and CM reference frames is given by:
\begin{equation}
	{E}_{ \text{proj} }^{ \text{(COSM)} } = {E}_{ \text{proj} }^{ \text{(CM)} } \frac{A}{ A - 1 } = \frac{ { \hbar }^{2} { \left( {k}_{ \text{proj} }^{ \text{(CM)} } \right) }^{2} }{ 2 {m}_{p} } \frac{A}{ A - 1 }
	\label{eq_rad_cap_p_energy_COSM}
\end{equation}
where ${ {k}_{ \text{proj} }^{ \text{(CM)} } }$ is the linear momentum of the projectile. Energy conservation implies that the final energy is:
${E}_{i}^{ \text{(COSM)} } = {E}_{f}^{ \text{(COSM)} } + {E}_{ \gamma }
\label{eq_rad_cap_tot_final_energy}$,
where ${ {E}_{f}^{ \text{(COSM)} } }$ is the compound system binding energy in the COSM frame of reference, and ${ {E}_{ \gamma } = {k}_{ \gamma } \hbar c }$ is the photon energy which does not depend on the chosen reference frame.

Resonances in the spectrum of a composite ${ A }$-nucleon system correspond to the peaks in the radiative capture cross section at the CM energy:
${E}_{ \text{CM} } = {E}_{i}^{ ( A ) } [ \text{GSM-CC} ] - {E}_{0}^{ ( A - 1 ) } [ \text{GSM} ]
\label{eq_rad_cap_resonance_position}$.
Here ${ {E}_{i}^{ ( A ) } [ \text{GSM-CC} ] }$ is the GSM-CC energy of the resonance '${ i }$' in the nucleus ${ A }$, and ${ {E}_{0}^{ (A) } [ \text{GSM} ] }$ is the GSM ground state energy of the target nucleus ${ ( A - 1 ) }$.

The cross section for a final state of the total angular momentum ${ {J}_{f} }$ is:
\begin{equation}
	{ \sigma }_{ {J}_{f} } ( {E}_{ \text{CM} } ) = \intt_{0}^{ 2 \pi } d{ \varphi }_{ \gamma } \intt_{0}^{ \pi } \sin{ \theta }_{ \gamma } d{ \theta }_{ \gamma } \frac{ d{ \sigma }_{ {J}_{f} } ( {E}_{ \text{CM} } , { \theta }_{ \gamma } , { \varphi }_{ \gamma } ) }{ d{ \Omega }{ \gamma } }
	\label{eq_rad_cap_partial_cross_section}
\end{equation}
and the total cross section is thus:
\begin{equation}
	\sigma ( {E}_{ \text{CM} } ) = \sum_{ {J}_{f} } { \sigma }_{ {J}_{f} } ( {E}_{ \text{CM} } )
	\label{eq_rad_cap_total_cross_section}
\end{equation}
In practice, one often shows the astrophysical factor:
\begin{equation}
	S ( {E}_{ \text{CM} } ) = \sigma ( {E}_{ \text{CM} } ) {E}_{ \text{CM} } {e}^{ 2 \pi \eta }
	\label{eq_rad_cap_astrophysical_factor}
\end{equation}
which removes the exponential dependence of the cross section at low energies due to the Coulomb barrier.
${ \eta }$ in \eqref{eq_rad_cap_astrophysical_factor} is the Sommerfeld parameter: $\eta = (mZ_1Z_2)/(\hbar^2k)$, where $Z_1$ and $Z_2$ are the proton numbers of the projectile and target nuclei.

\subsection{Parameters of GSM calculations in $^7$Be and $^8$B}
\label{sec3b}

The model space in ${ {}^{7}\text{Be} }$ and ${ {}^{8}\text{B} }$ is limited by the core of ${ {}^{4}\text{He} }$. 
The core is described by a WS potential (see Table \ref{tab_WS_4He_core_for_7Be}) for each considered partial wave: ${ l = 0, 1 }$ and 2. The radius of the Coulomb potential is ${ {r}_{c} = 2.8 \, \unitsText{fm} }$.
\begin{table}[ht!]
	\centering
	\begin{tabular}{|c|c|c|}
		\hline
		Parameter & Protons & Neutrons \\
		\hline
		${ a }$ 				& 0.65 fm 	& 0.65 fm \\
		${ {R}_{0} }$ 				& 2.0 fm 	& 2.0 fm \\
		${ {V}_{ \text{o} } ( l = 0 ) }$ 	& 61.5 MeV 	& 70.6735 MeV \\
		${ {V}_{ \text{so} } ( l = 0 ) }$ 	& 0 MeV 	& 0 MeV \\
		${ {V}_{ \text{o} } ( l = 1 ) }$ 	& 44.3967 MeV   & 70.6734 MeV \\
		${ {V}_{ \text{so} } ( l = 1 ) }$ 	& 7.80188 MeV   & 7.86276 MeV \\
		${ {V}_{ \text{o} } ( l = 2 ) }$ 	& 44.3967 MeV   & 0 MeV \\
		${ {V}_{ \text{so} } ( l = 2 ) }$ 	& 7.80188 MeV   & 0 MeV \\
		\hline
	\end{tabular}
	\caption{Parameters of the WS potential of ${ {}^{4}\text{He} }$ core used in the GSM and GSM-CC description of ${ {}^{7}\text{Be} }$ and ${ {}^{8}\text{B} }$.}
	\label{tab_WS_4He_core_for_7Be}
\end{table}
To determine Berggren ensemble, one calculates first the single-particle bound and resonance states of the basis generating
WS potential for all chosen partial waves $(l,j)$. Then, for each $(l,j)$, one selects the contour ${\cal L}^+_{l j}$ in
a fourth quadrant of the complex $k$-plane. All $(l,j)$-scattering states in this ensemble belong to ${\cal L}^+_{l j}$ . The 
precise form of the contour is unimportant providing that all selected single-particle resonances for a given $(l,j)$ lie between this contour and the real $k$-axis for ${\cal R}(k) > 0$. For each $(l,j)$, the set of all resonant states and scattering states on ${\cal L}^+_{l j}$ forms a complete single-particle basis. 

In the present case, valence nucleons can occupy the ${ 0{p}_{ 3/2 } }$ and ${ 0{p}_{ 1/2 } }$ discrete single-particle states and several non-resonant single-particle continuum states on discretized contours: ${ {\cal L}_{ {s}_{ 1/2 } }^{+} }$, ${ {\cal L}_{ {p}_{ 1/2 } }^{+} }$, ${ {\cal L}_{ {p}_{ 3/2 } }^{+} }$, ${ {\cal L}_{ {d}_{ 3/2 } }^{+} }$ and ${ {\cal L}_{ {d}_{ 5/2 } }^{+} }$. Each contour consists of three segments joining the points: $k_{\rm min} = 0.0$, ${ {k}_{ \text{peak} } = 0.15 - i 0.14 \, \unitsText{fm}^{-1} }$, ${ {k}_{ \text{middle} } = 0.3 \, \unitsText{fm}^{-1} }$ and ${ {k}_{ \text{max} } = 2.0 \, \unitsText{fm}^{-1} }$, and each segment is discretized with 10 points. Hence, GSM and GSM-CC calculations are done in 152 shells: 31 ${ {p}_{ 3/2 } }$ and ${ {p}_{ 1/2 } }$ shells, and 30 ${ {s}_{ 1/2 } }$, ${ {d}_{ 3/2 } }$ and ${ {d}_{ 5/2 } }$ shells.
The GSM basis is truncated so as to reduce the size of the GSM Hamiltonian matrix. For this, the occupation of  ${ {p}_{ 3/2 } }$ and ${ {p}_{ 1/2 } }$ scattering states in basis Slater determinants is limited to two particles,
while the occupation of  ${ {s}_{ 1/2 } }$, ${ {d}_{ 5/2 } }$ and ${ {d}_{ 3/2 } }$ scattering states is limited to one particle only. 
The latter truncation is justified by the fact that GSM target states virtually only consists of ${ {p}_{ 3/2 } }$ and ${ {p}_{ 1/2 } }$ states,
${ {s}_{ 1/2 } }$, ${ {d}_{ 5/2 } }$ and ${ {d}_{ 3/2 } }$  states occurring only in the partial wave decomposition of the proton or neutron projectile.

Parameters of the Hamiltonian, which were adjusted to reproduce binding energies of low-lying states in $^7$Be and $^8$B, are given in Table \ref{tab_FHT_7Be_8B}.
\begin{table}[ht!]
	\centering
	\begin{tabular}{|c|c|}
		\hline
		Parameter & Value [MeV] \\
		\hline 
		${ {\cal V}_{ \text{t,t} }^{\rm C} }$ 	& 4.00906 \\
		${ {\cal V}_{ \text{s,t} }^{\rm C} }$ 	& -3.22579 \\
		${ {\cal V}_{ \text{s,s} }^{\rm C} }$ 	& 2.22077 \\
		${ {\cal V}_{ \text{t,s} }^{\rm C} }$ 	& -9.51008 \\
		${ {\cal V}_{ \text{t,t} }^{\rm SO} }$ 	& -1448.32 \\
		${ {\cal V}_{ \text{s,t} }^{\rm SO} }$ 	& 0 \\
		${ {\cal V}_{ \text{t,t} }^{\rm T} }$ 	& 15.3946 \\
		${ {\cal V}_{ \text{s,t} }^{\rm T} }$ 	& -15.4834 \\
		\hline
	\end{tabular}
	\caption{Parameters of the FHT interaction in GSM and GSM-CC calculations in ${ {}^{7}\text{Be} }$ and ${ {}^{8}\text{B} }$. The superscripts ``C'', ``SO'' and ``T'' stand for 'central', 'spin-orbit' and 'tensor', respectively, and the indices ``s'' and ``t'' stand for 'singlet' and 'triplet'.}
	\label{tab_FHT_7Be_8B}
\end{table}
In GSM calculations, the ground state of ${ {}^{7}\text{Be} }$ is bound with respect to ${ {}^{4}\text{He} }$ by ${ 9.378 \, \unitsText{MeV} }$, close to the experimental value ${ {E}_{ \text{exp} } = 9.304 \, \unitsText{MeV} }$.
Reaction channels in GSM-CC calculations are obtained by the coupling of the ground state ${ { 3/2 }^{-} }$ and the first excited state ${ { 1/2 }^{-} }$ of ${ {}^{7}\text{Be} }$ with the proton partial waves: ${ {s}_{ 1/2 } }$, ${ {p}_{ 1/2 } }$, ${ {p}_{ 3/2 } }$, ${ {d}_{ 3/2 } }$ and ${ {d}_{ 5/2 } }$.  

Discrete states of a composite system $^8$B are ${ {2}_{1}^{+} }$ bound state, and ${ {1}_{1}^{+} }$, ${ {3}_{1}^{+} }$, $1_2^+$ resonances. Missing reaction channels in GSM-CC lead to a small difference between GSM and GSM-CC energies for these states. To correct this deficiency, the channel-channel coupling potentials ${ {V}_{ c , c' } }$ in GSM-CC have been adjusted for each considered state of ${ {}^{8}\text{B} }$.
The new potentials are: ${ \tilde{V}_{ c , c' } = c ( {J}^{ \pi } ) {V}_{ c , c' } }$, where the multiplicative corrective factors are:
${ c ( {2}_{1}^{+} ) = 1.0133 }$, ${ c ( {1}_{1}^{+} ) = 1.0602 }$, and ${ c ( {3}_{1}^{+} ) = 1.0233 }$.

\subsection{The astrophysical $S$-factor for $^7$Be$(p,\gamma)$$^8$B reaction}
\label{sec3c}

The description of electromagnetic transitions requires effective charges for proton and neutron.
For E1 transitions, the standard values are \cite{hornyak75_b130}:
\begin{equation}
	{e}_{ \text{eff} }^{p} = e \left( 1 - \frac{Z}{A} \right) \ ; \qquad
	{e}_{ \text{eff} }^{n} = - e \frac{Z}{A}
	\label{eq_rad_cap_eff_charges_E1}
\end{equation}
where ${ Z }$ and ${ A }$ are the proton number and the total number of nucleons, respectively.
The standard values for E2 transitions are:
\begin{equation}
	{e}_{ \text{eff} }^{p} = e \left( 1 - \frac{Z}{A} + \frac{Z}{ {A}^{2} } \right) \ ; \qquad
	{e}_{ \text{eff} }^{n} = - e \frac{Z}{ {A}^{2} }
	\label{eq_rad_cap_eff_charges_E2}
\end{equation}
There are no effective charges for M1 transitions. In the present work, we use these standard values for E1 and E2 effective charges. One should keep in mind however, that the effective charges extracted experimentally show often significant deviations from the standard values \cite{bohr98_b27}.

\begin{figure}[htb]
	\includegraphics[width=0.85\linewidth]{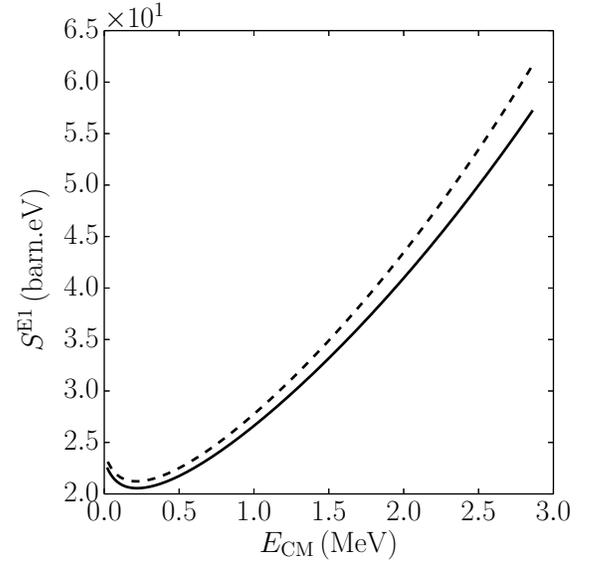}
	\caption{Plot of the E1 astrophysical factor for the ${ {}^{7}\text{Be} ( p , \gamma ) {}^{8}\text{B} }$ reaction. The solid line represents the exact, fully antisymmetrized GSM-CC calculation with both the ground state $J^{\pi}=3/2_1^-$ and the first excited state $J^{\pi}=1/2_1^-$ of ${ {}^{7}\text{Be}}$ target included. The dashed line shows results of the GSM-CC calculations if the first excited state of ${ {}^{7}\text{Be}}$ is omitted. For more details, see the description in the text.}
	\label{fig_1}
\end{figure}
\begin{figure}[htb]	
	\includegraphics[width=0.85\linewidth]{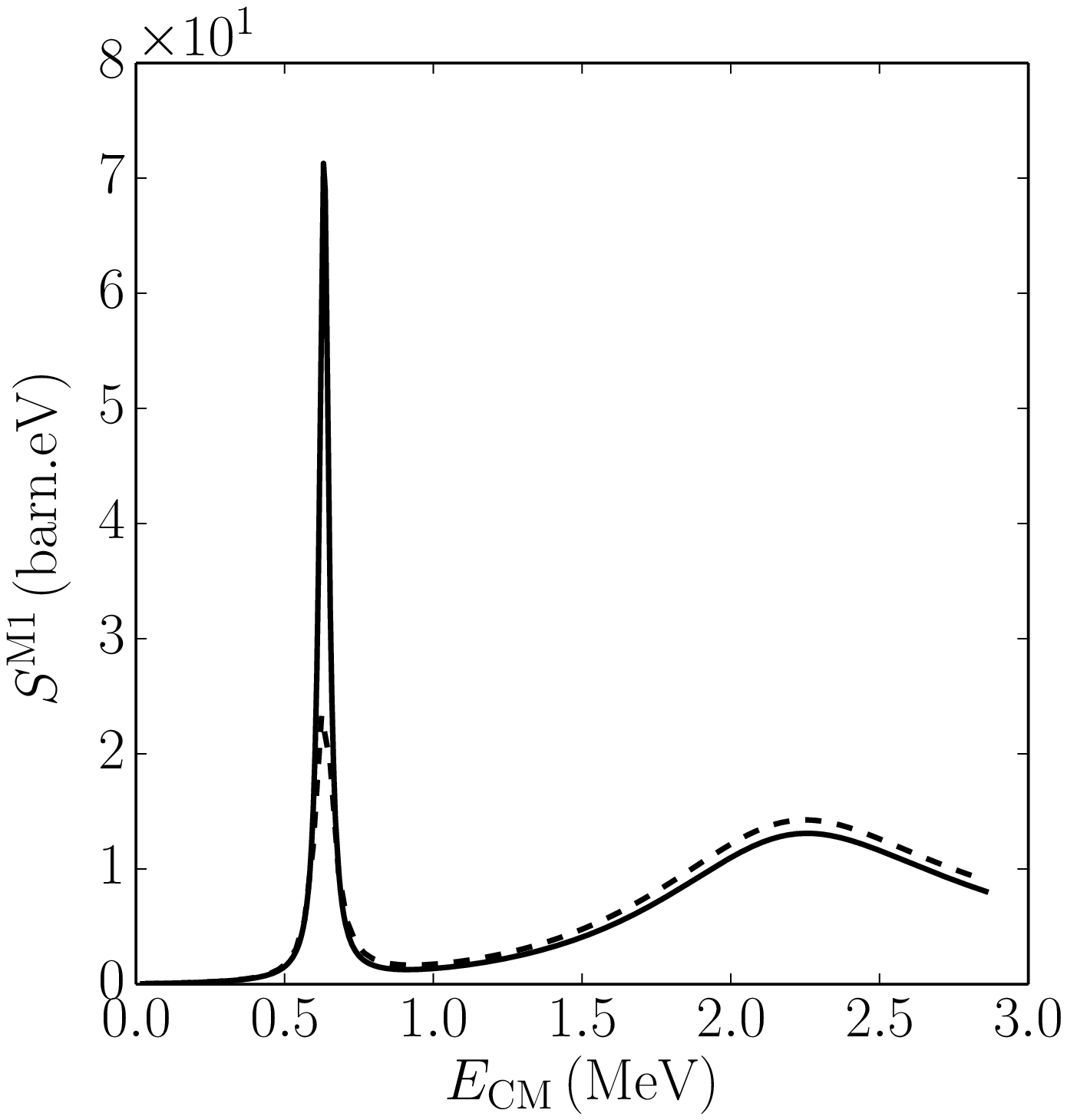}
	\caption{The same as in Fig.\ref{fig_1} but for the M1 transitions. The two peaks correspond to the ${ {1}_{1}^{+} }$ and ${ {3}_{1}^{+} }$ resonances of ${ {}^{8}\text{B} }$.}
	\label{fig_2}
\end{figure}

\begin{figure}[htb]
	\includegraphics[width=0.85\linewidth]{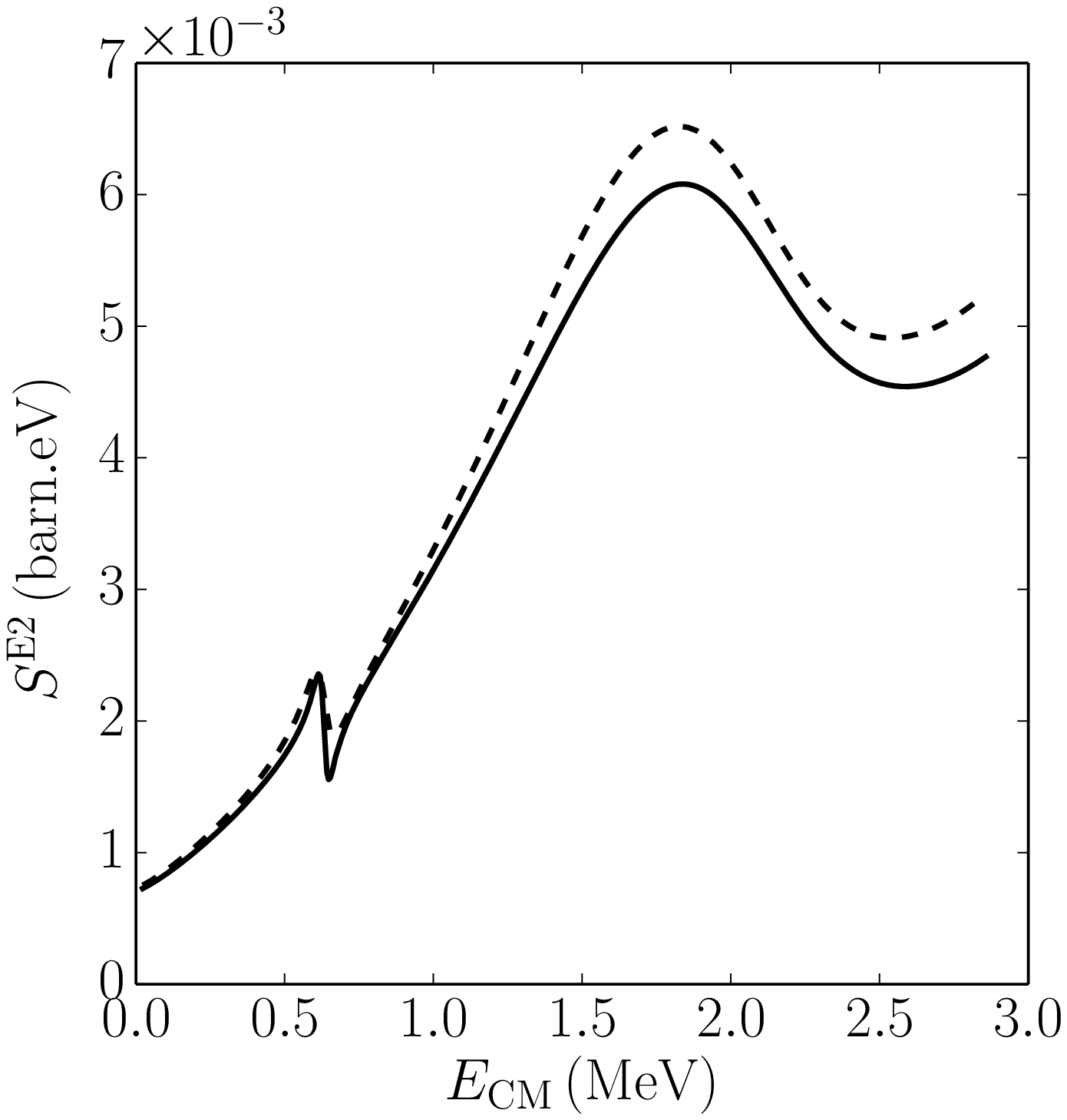}
	\caption{The same as in Fig.\ref{fig_1} but for the E2 transitions. The two peaks correspond to the ${ {1}_{1}^{+} }$ and ${ {1}_{2}^{+} }$ resonances of ${ {}^{8}\text{B} }$.}
		\label{fig_3}
\end{figure}
\begin{figure}[htb]
	\includegraphics[width=0.85\linewidth]{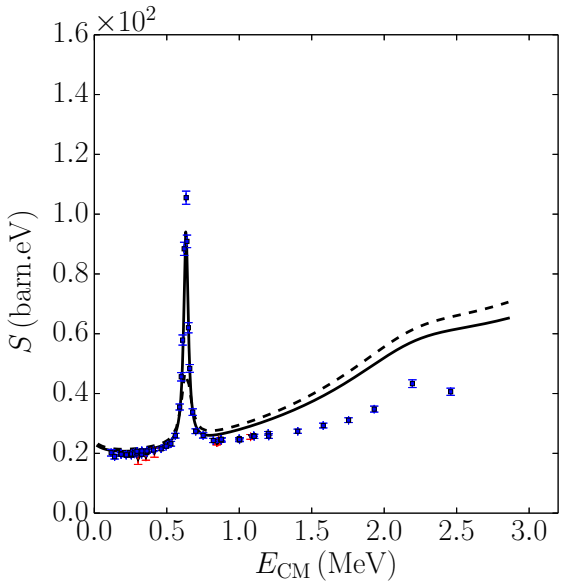}
	\caption{Plot of the total astrophysical factor for the ${ {}^{7}\text{Be} ( p , \gamma ) {}^{8}\text{B} }$ reaction. Data are taken from Refs.\cite{baby03_900} and \cite{junghans10_904}. The solid line represents the exact, fully antisymmetrized GSM-CC calculation including both the ground state $J^{\pi}=3/2_1^-$ and the first excited state $J^{\pi}=1/2_1^-$ of ${ {}^{7}\text{Be}}$ target. Calculations neglecting the first excited state of the target are shown with the dashed line. For more details see the description in the text.}
	\label{fig_4}
\end{figure}
Proton separation energy in the ground state of ${ {}^{7}\text{Be} }$ is $S_{\rm p}={ 5.6 \, \unitsText{MeV} }$. The final nucleus ${ {}^{8}\text{B} }$ has one weakly bound state ${2}_{1}^{+}$ below the proton emission threshold. Experimental proton separation energy in this state ${ {S}_{ \text{p} } = 0.1375 \, \unitsText{MeV} }$ agrees well with the calculated value ${ {S}_{ \text{p} }^{ ( \text{th} ) } = 0.137 \, \unitsText{MeV} }$. The ${ {1}_{1}^{+} }$ and ${ {3}_{1}^{+} }$ resonance peaks should be seen in M1 transitions. The ${ {1}_{1}^{+} }$ resonance could also be seen in E2 transitions.

All relevant E1, M1, E2 transitions from the initial continuum states ($J_i = 1^+, 2^+, 3^+$) in $^8$B to the final bound state $J_f = 2^+$ state have been included. Figs. \ref{fig_1}-\ref{fig_3} show the separate contributions to the total $S$-factor in ${ {}^{7}\text{Be} ( p , \gamma ) {}^{8}\text{B} }$ reaction: $S^{\rm E1}$ for E1 transitions (Fig. \ref{fig_1}), $S^{\rm M1}$ for M1 transitions (Fig. \ref{fig_2}), and $S^{\rm E2}$ for E2 transitions. The solid lines in Figs. \ref{fig_1}-\ref{fig_3} show results of the fully antisymmetrized GSM-CC calculations with both ground and first excited states of the $^7$Be target included. The dashed line in these figures correspond to GSM-CC calculations neglecting the $1/2^-$ first excited state in $^7$Be. 

There is no resonant contribution in E1 transitions. Including the first excited state of the target lowers $S^{\rm E1}$ by less than $\sim 5 \%$ for $E_{\rm CM}<2.5$ MeV. On the contrary, the M1 contribution to the $S$-factor increases significantly in the region of $1_1^+$ resonance if the the excited state of the target is included (see Fig. \ref{fig_2}). One can see ${ {1}_{1}^{+} }$ and ${ {3}_{1}^{+} }$ resonances of ${ {}^{8}\text{B} }$ at ${ {E}_{ \text{CM} } = 0.79 \, \unitsText{MeV} }$ and ${ {E}_{ \text{CM} } = 2.34 \, \unitsText{MeV} }$, respectively. These resonances are observed experimentally at ${ {E}_{ \text{CM} } = 0.632 \, \unitsText{MeV} }$ and ${ {E}_{ \text{CM} } = 2.182 \, \unitsText{MeV} }$, respectively. 
The E2 transitions contribute little to the $S$-factor. $S^{\rm E2}$ is $\sim$10$^{-3}$ smaller than $S^{\rm E1}$ and $S^{\rm M1}$ and increases by less than $\sim 10 \%$ for CM energies in the region of $1_1^+$ and $1_2^+$ resonances. The resonance $1_2^+$ has not yet been seen experimentally.

The calculated total $S$-factor is compared with the experimental data \cite{baby03_900,junghans10_904} in Fig. \ref{fig_4}. Below $E_{\rm CM} = 1$ MeV, the agreement with the data  is good if both the ground state of $^7$Be and its first excited state are included. The value of the $S$-factor at zero energy, $S^{\rm GSM-CC}(0)$, is 23.214 b$\cdot$eV and the slope, ${\partial S}/{\partial E_{\rm CM}}|_{E_{\rm CM}=0}$, is 37.921b. The accepted experimental value of the $S$-factor is 20.9$\pm$0.6 b$\cdot$eV, slightly below the GSM-CC results.

At higher energies, GSM-CC results overshoot the experimental data. This feature could be due to the absence of higher lying discrete and continuum states of $^7$Be target in the channel basis. Indeed, in the present case, GSM and GSM-CC calculations with uncorrected channel-channel coupling potentials $V_{c,c'}$ do not give the same spectra and binding energies of $^7$Be and $^8$B and the small multiplicative correction factors are necessary.

\begin{figure}[htb]
	\includegraphics[width=0.85\linewidth]{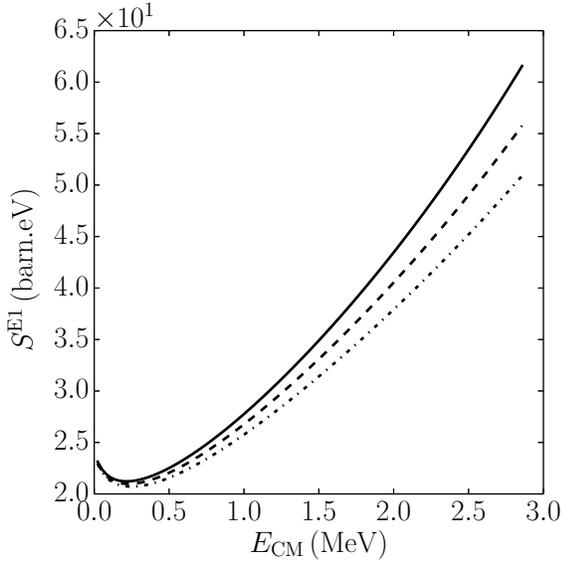}
	\caption{Plot of the E1 astrophysical factor for the ${ {}^{7}\text{Be} ( p , \gamma ) {}^{8}\text{B} }$ reaction. The solid line represents the exact, fully antisymmetrized GSM-CC calculation. The calculations in the long wavelength approximation are represented by the dashed and dotted lines in the fully antisymmetrized and non-antisymmetrized cases, respectively. For more details, see the description in the text.}
	\label{fig_5}
\end{figure}
\begin{figure}[htb]
	\includegraphics[width=0.85\linewidth]{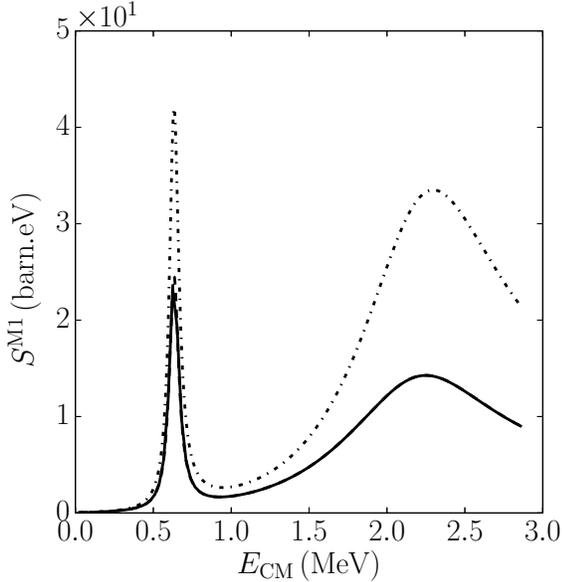}
	\caption{The same as in Fig.\ref{fig_5} but for the M1 transitions. The two peaks correspond to the ${ {1}_{1}^{+} }$ and ${ {3}_{1}^{+} }$ resonances of ${ {}^{8}\text{B} }$.}
	\label{fig_6}
\end{figure}
The long wavelength approximation simplifies the calculation of matrix elements of the electromagnetic transitions. The quality of this approximation and the role of the antisymmetry of initial and final states is tested in Figs. \ref{fig_5} and \ref{fig_6}. Only the ground state of $^7$Be is taken into account. To correct GSM-CC calculations for the missing channels in this case, the channel-channel coupling potentials ${ {V}_{ c , c' } }$ have been slightly corrected: ${ \tilde{V}_{ c , c' } = c ( {J}^{ \pi } ) {V}_{ c , c' } }$, and the  multiplicative corrective factors are: ${ c ( {2}_{1}^{+} ) = 1.0122 }$, ${ c ( {1}_{1}^{+} ) = 1.0668 }$, and ${ c ( {3}_{1}^{+} ) = 1.0225 }$.

At low energies ($E_{\rm CM} < 1.5$ MeV), both the long wavelength approximation and the antisymmetrization in the calculation of E1 transition matrix elements does not change results significantly (see Fig. \ref{fig_5}). Both approximations become worse at higher energies but even at $E_{\rm CM} = 2.5$ MeV the error is only $\sim 10\%$. 

The astrophysical factor for M1 transitions is shown in Fig. \ref{fig_6}.
The antisymmetrization of the initial and final states lowers the value of $S^{\rm M1}$ by a factor ${ \sim 2 }$  at the resonance peaks. The long wavelength approximation does not change $S^{\rm M1}$.

\section{Results of GSM-CC calculations for the $^7$Li(n,$\gamma$)$^8$Li reaction} 
\label{sec4}

\subsection{Parameters of GSM calculations in $^7$Li and $^8$Li}
\label{sec4a}

${ {}^{7}\text{Li} ( n , \gamma ) {}^{8}\text{Li} }$ is the mirror reaction of ${ {}^{7}\text{Be} ( p , \gamma ) {}^{8}\text{B} }$ and will be described in the same model space.The WS potential of ${ {}^{4}\text{He} }$ core is given in Table \ref{tab_WS_4He_core_for_7Li}. The radius of the Coulomb potential is ${ {r}_{c} = 2.8 \, \unitsText{fm} }$.
\begin{table}[ht!]
	\centering
	\begin{tabular}{|c|c|c|}
		\hline
		Parameter & Protons & Neutrons \\
		\hline
		${ a }$ 				& 0.65 fm 	& 0.65 fm \\
		${ {R}_{0} }$ 				& 2.0 fm 	& 2.0 fm \\
		${ {V}_{ \text{o} } ( l = 0 ) }$ 	& 71.0752 MeV	& 43.6438 MeV \\
		${ {V}_{ \text{so} } ( l = 0 ) }$ 	& 0 MeV 	& 0 MeV \\
		${ {V}_{ \text{o} } ( l = 1 ) }$ 	& 71.0752 MeV   & 43.6438 MeV \\
		${ {V}_{ \text{so} } ( l = 1 ) }$ 	& 7.90622 MeV   & 7.84517 MeV \\
		${ {V}_{ \text{o} } ( l = 2 ) }$ 	& 0 MeV 	& 43.6438 MeV \\
		${ {V}_{ \text{so} } ( l = 2 ) }$ 	& 0 MeV 	& 0 MeV \\
		\hline
	\end{tabular}
	\caption{Parameters of the WS potential of   ${ {}^{4}\text{He} }$ core used in the GSM and GSM-CC description of ${ {}^{7}\text{Li} }$ and ${ {}^{8}\text{Li} }$ nuclei.}
	\label{tab_WS_4He_core_for_7Li}
\end{table}
Valence nucleons occupy ${ 0{p}_{ 3/2 } }$ and ${ 0{p}_{ 1/2 } }$ discrete single-particle states and non-resonant single-particle continuum states on discretized contours: ${ {\cal L}_{ {s}_{ 1/2 } }^{+} }$, ${ {\cal L}_{ {p}_{ 1/2 } }^{+} }$, ${ {\cal L}_{ {p}_{ 3/2 } }^{+} }$, ${ {\cal L}_{ {d}_{ 3/2 } }^{+} }$ and ${ {\cal L}_{ {d}_{ 5/2 } }^{+} }$. Each contour consists of three segments joining the points: $k_{\rm min} = 0.0$, ${ {k}_{ \text{peak} } = 0.15 - i 0.14 \, \unitsText{fm}^{-1} }$, ${ {k}_{ \text{middle} } = 0.3 \, \unitsText{fm}^{-1} }$ and ${ {k}_{ \text{max} } = 2.0 \, \unitsText{fm}^{-1} }$, and each segment is discretized by 10 points. 

Parameters of the FHT Hamiltonian in ${ {}^{7}\text{Li} }$ and ${ {}^{8}\text{Li} }$ are given in Table \ref{tab_FHT_7Li_8Li}.
\begin{table}[ht!]
	\centering
	\begin{tabular}{|c|c|}
		\hline
		Parameter & Value [MeV] \\
		\hline
		${ {\cal V}_{ \text{t,t} }^{\rm C} }$ 	& 4.03185 \\
		${ {\cal V}_{ \text{s,t} }^{\rm C} }$ 	& -4.95286 \\
		${ {\cal V}_{ \text{s,s} }^{\rm C} }$ 	& 2.23361 \\
		${ {\cal V}_{ \text{t,s} }^{\rm C} }$ 	& -7.63465 \\
		${ {\cal V}_{ \text{t,t} }^{\rm SO} }$ 	& -1456.55 \\
		${ {\cal V}_{ \text{s,t} }^{\rm SO} }$ 	& 0 \\
		${ {\cal V}_{ \text{t,t} }^{\rm T} }$ 	& 15.4822 \\
		${ {\cal V}_{ \text{s,t} }^{\rm T} }$ 	& -15.5716 \\
		\hline
	\end{tabular}
	\caption{Parameters of the FHT interaction for GSM and GSM-CC calculations in ${ {}^{7}\text{Li} }$ and ${ {}^{8}\text{Li} }$. For more details, see the caption of Table \ref{tab_FHT_7Be_8B}.}
	\label{tab_FHT_7Li_8Li}
\end{table}
\setlength{\parskip}{0pt}
In GSM, the ground state of ${ {}^{7}\text{Li} }$ is bound by ${ 11.228 \, \unitsText{MeV} }$ with respect to ${ {}^{4}\text{He} }$, \textit{i.e.} close to the experimental value (${ {E}_{ \text{exp} } = 10.948 \, \unitsText{MeV} }$).
Reaction channels are obtained by the coupling of the ground state ${ { 3/2 }^{-} }$ and the first excited state ${ { 1/2 }^{-} }$ of ${ {}^{7}\text{Li} }$ with the proton partial waves: ${ {s}_{ 1/2 } }$, ${ {p}_{ 1/2 } }$, ${ {p}_{ 3/2 } }$, ${ {d}_{ 3/2 } }$ and ${ {d}_{ 5/2 } }$.

Discrete states of a composite system ${ {}^{8}\text{Li} }$ are ${ {2}_{1}^{+} }$, ${ {1}_{1}^{+} }$ bound states, and ${ {3}_{1}^{+} }$ resonance. To correct for missing reaction channels in GSM-CC calculations, the channel-channel coupling potentials ${ {V}_{ c , c' } }$ have been modified and new potentials are: ${ \tilde{V}_{ c , c' } = c ( {J}^{ \pi } ) {V}_{ c , c' } }$, with ${ c ( {2}_{1}^{+} ) = 1.03705 }$, ${ c ( {1}_{1}^{+} ) = 1.04805 }$ and ${ c ( {3}_{1}^{+} ) = 1.03205 }$.

\subsection{$^7$Li$(n,\gamma)$$^8$Li cross section}
\label{sec4b}

Neutron separation energy in the ground state of ${ {}^{7}\text{Li} }$ is $S_{\rm n} = { 7.25 \, \unitsText{MeV} }$. The final nucleus ${ {}^{8}\text{Li} }$ has two bound states ${ {J}^{ \pi } = {2}_{1}^{+} }$ and ${ {1}_{1}^{+} }$ below the neutron emission threshold. Neutron separation energy from the ground state and excited states are ${ {S}_{ \text{n} } = 2.03262 \, \unitsText{MeV} }$ and ${ {S}_{ \text{n} } = ... \, \unitsText{MeV} }$, respectively. The calculated neutron separation energies in these two states are ${ {S}_{ \text{n} }^{ ( \text{th} ) } = 2.032 \, \unitsText{MeV} }$ and ${ {S}_{ \text{n} }^{ ( \text{th} ) } = ... \, \unitsText{MeV} }$, in good agreement with the experimental data. The ${ {3}_{1}^{+} }$ resonance peak can be seen in M1 and E2 transitions.

\vskip 0.2truecm
\begin{figure}[htb]
	\includegraphics[width=0.85\linewidth]{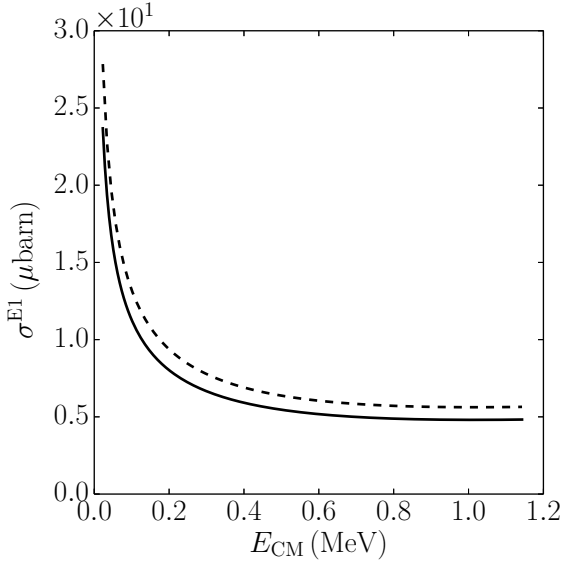}
	\caption{The same as in Fig.\ref{fig_1} but for the reaction ${ {}^{7}\text{Li} ( p , \gamma ) {}^{8}\text{Li} }$ reaction.  For more details, see the description in the text.}
	\label{fig_7}
\end{figure}
\begin{figure}[htb]
	\includegraphics[width=0.85\linewidth]{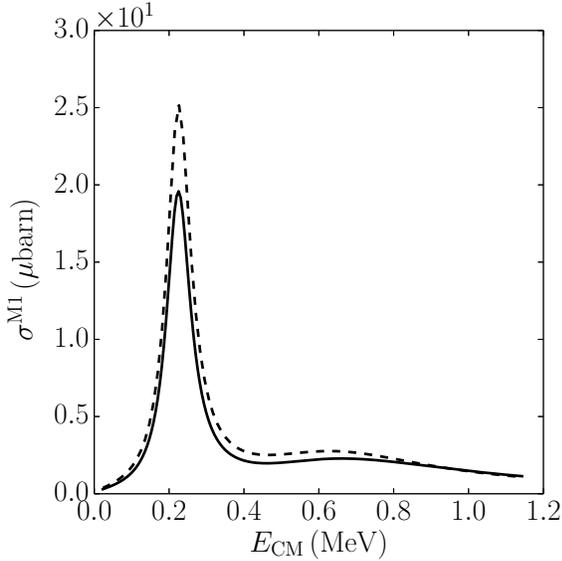}
	\caption{The same as in Fig.\ref{fig_1} but for M1 transitions in ${ {}^{7}\text{Li} ( p , \gamma ) {}^{8}\text{Li} }$ reaction. The peak corresponds to the ${ {3}_{1}^{+} }$ resonance in ${ {}^{8}\text{Li} }$.}
	\label{fig_8}
\end{figure}

\begin{figure}[htb]
	\includegraphics[width=0.85\linewidth]{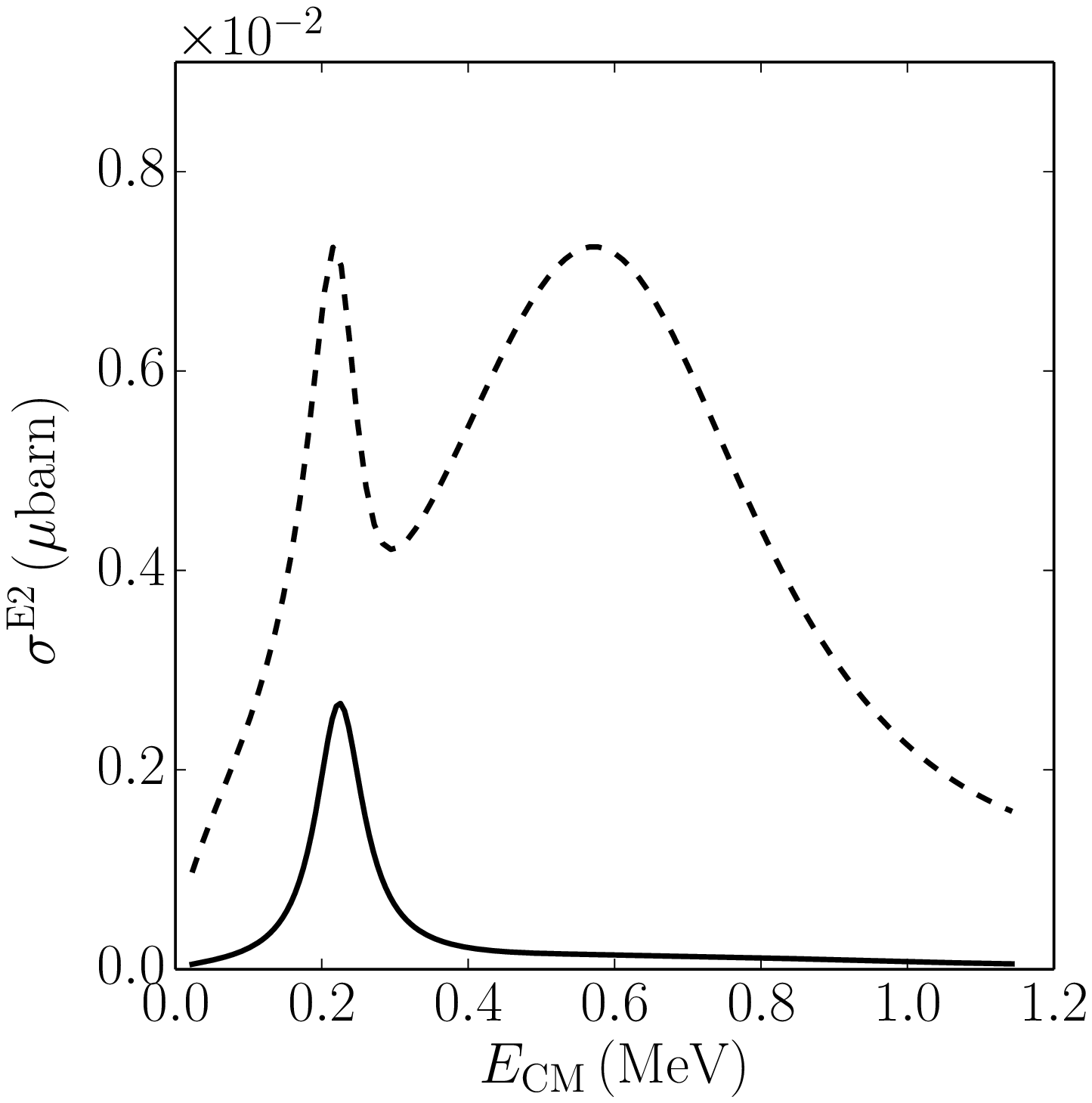}
	\caption{The same as in Fig.\ref{fig_8} but for the E2 transitions. The two peaks correspond to the ${ {3}_{1}^{+} }$ and ${ {1}_{2}^{+} }$ resonances of ${ {}^{8}\text{Li} }$.}
	\label{fig_9}
\end{figure}
\begin{figure}[htb]
	\includegraphics[width=0.85\linewidth]{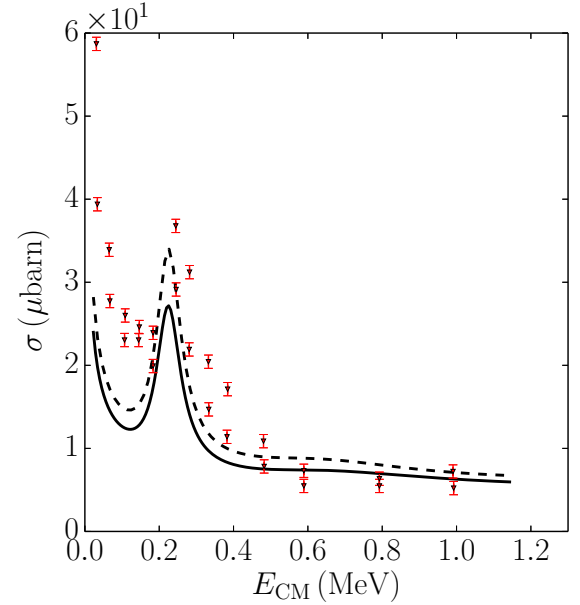}
	\caption{Plot of the total cross section for the ${ {}^{7}\text{Li} ( n , \gamma ) {}^{8}\text{Li} }$ reaction. Data are taken from Ref.\cite{imhof59_1006}. The solid line represents the exact, fully antisymmetrized GSM-CC calculation. Calculations in the long wavelength approximation are represented by the dashed and dotted lines in the antisymmetrized and non-antisymmetrized cases, respectively. For more details see the description in the text.}
	\label{fig_10}
\end{figure}
Figs. \ref{fig_7}-
\ref{fig_9} show the E1, M1 and E2 cross sections for ${ {}^{7}\text{Li} ( n , \gamma ) {}^{8}\text{Li} }$ reaction. 
The solid lines in Figs. \ref{fig_7}-
\ref{fig_9} show results of the fully antisymmetrized GSM-CC calculations with both ground and first excited states of the $^7$Li included. The dashed line in these figures correspond to GSM-CC calculations neglecting the $1/2^-$ first excited state in $^7$Li. Including the first excited state of the target lowers E1 contribution to the neutron radiative capture cross-section by $\sim 20 \%$ for $E_{\rm CM}<1$ MeV. 

The M1 contribution to the cross-section increases by $\sim 25 \%$ in the region of $3_1^+$ resonance if the excited state of the target is included (see Fig. \ref{fig_8}). One can see that the calculated ${ {3}_{1}^{+} }$ resonance is at the experimental value of energy: ${ {E}_{ \text{CM} } = 0.223 \, \unitsText{MeV} }$.

E2 transitions contribute very little to the neutron radiative capture cross section. The E2 contribution is three orders of magnitude smaller than E1 and M1 contributions. The role of the excited state of the target is very important. It increases the contribution from E2 transitions by a factor $\sim 3$ in the region of $3_1^+$ resonance. At the $1_2^+$ resonance, the excited state enhances the E2 contribution by about one order of magnitude. The calculated energy of this resonance is lower than seen experimentally.

The total neutron radiative capture cross section is compared with the experimental data \cite{imhof59_1006} in Fig. \ref{fig_10}. GSM-CC calculation underestimates the data of Imhof et al \cite{imhof59_1006}. The extrapolation of the calculated neutron radiative capture cross section at low ${ {E}_{ \text{CM} } }$ is done using the expansion:
\begin{equation}
	\sigma ( {E}_{ \text{CM} } ) = \frac{4.541}{ \sqrt{ {E}_{ \text{CM} } } } - 2.360 + 3.387 \sqrt{ {E}_{ \text{CM} } }
	\label{eq_fit_cross_section_s_wave_neutron_capture}
\end{equation}
which yields: ${ { \sigma }^{ \text{(GSM-CC)} } = 25.41 \, \unitsMath{\mu}\unitsText{barn} }$ at $E_{\rm CM} = 25$ keV.

\begin{figure}[htb]
	\includegraphics[width=0.85\linewidth]{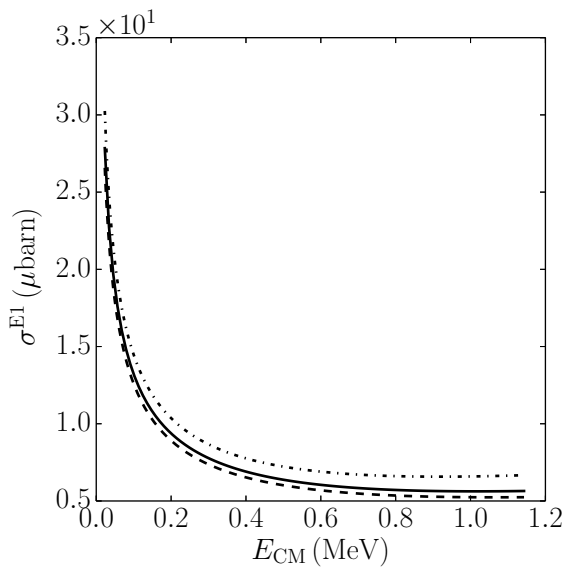}
	\caption{The same as in Fig.\ref{fig_5} but for ${ {}^{7}\text{Li} ( p , \gamma ) {}^{8}\text{Li} }$ reaction. For more details, see the description in the text.}
	\label{fig_11}
\end{figure}
\begin{figure}[htb]
	\includegraphics[width=0.85\linewidth]{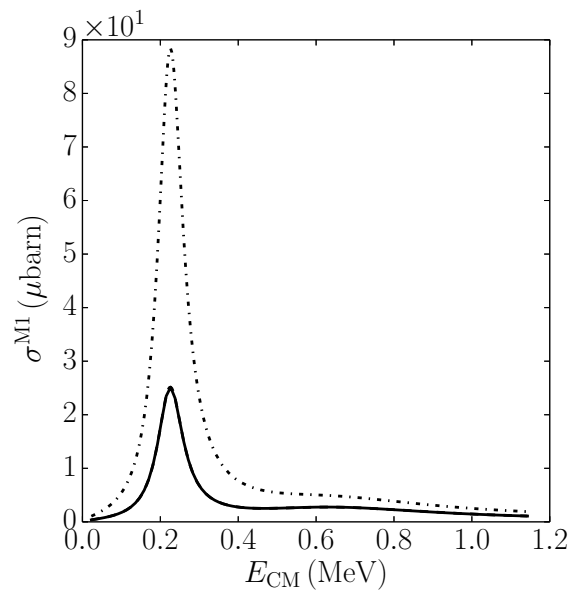}
	\caption{The same as in Fig.\ref{fig_11} but for the M1 transitions. The peak corresponds to the ${ {3}_{1}^{+} }$ resonance in ${ {}^{8}\text{Li} }$.}
	\label{fig_12}
\end{figure}
The long wavelength approximation and the role of the antisymmetry of initial and final states in the calculation of matrix elements of the electromagnetic transitions is tested in Figs. \ref{fig_11}, \ref{fig_12}. Only the ground state of $^7$Li is taken into account in these tests. 
To correct GSM-CC calculations for the missing channels in this case, the channel-channel coupling potentials ${ {V}_{ c , c' } }$ have been corrected: ${ \tilde{V}_{ c , c' } = c ( {J}^{ \pi } ) {V}_{ c , c' } }$, and the  multiplicative corrective factors are: ${ c ( {2}_{1}^{+} ) = 1.038 }$, ${ c ( {1}_{1}^{+} ) = 1.0594 }$, and ${ c ( {3}_{1}^{+} ) = 1.032 }$.

At low energies ($E_{\rm CM} < 1.2$ MeV), both the long wavelength approximation and the antisymmetry of initial and final states in the calculation of E1 transition matrix elements does not change results significantly (see Fig. \ref{fig_11}). Also M1 transition matrix elements are insensible to the long wavelength approximation (see Fig. \ref{fig_12}). On the contrary, the antisymmetrization is essential, decreasing the M1 contribution to the neutron radiative capture cross section by a factor $\sim 4$ in the region of $3_1^+$ resonance.

\section{Conclusions}
\label{sec5}

The GSM in the coupled-channel representation opens a possibility for the unified description of low-energy nuclear structure and reactions using the same Hamiltonian. While both GSM and GSM-CC can describe energies, widths and wave functions of the many-body states, the GSM-CC can in addition yield reaction cross-sections. Combined application of GSM and GSM-CC to describe energies of resonant states allows to test the exactitude of calculated cross-sections for a given many-body Hamiltonian.

In this work, we have presented in details  the GSM in the coupled channel representation and applied it for the description of the low-energy proton and neutron radiative capture processes on mirror targets $^7$Be and $^7$Li, respectively. The interaction between valence nucleons in this calculation was modelled by the finite-range two-body FHT interaction. 

The convergence of GSM-CC calculations has been checked by comparing GSM and GSM-CC results for $^{8}$B and $^8$Li states. In a given single-particle model space, the GSM-CC calculation with the reaction channels which are constructed using selected many-body states of the target nucleus ($^7$Be or $^7$Li in our case), can be considered reliable if the GSM-CC eigenvalues for a combined system ($^8$B or $^8$Li in our case) approximate well results of a direct diagonalization of the GSM Hamiltonian matrix in the same single-particle model space. In such a case, the configuration mixing in GSM-CC and GSM wave functions are equivalent and one does not need to include additional states of the target nucleus to reach the many-body completeness in GSM-CC calculation. Only in this case, the unified description of nuclear structure and reactions with the same many-body Hamiltonian and the same model space is reached. In the studied case, the GSM and GSM-CC spectra were close but not identical so the small renormalization of the channel-channel coupling potentials was necessary to compensate for the missing channels made of the higher lying discrete and/or continuum states of the target.

There are two important aspects in this GSM-CC calculations which have been studied carefully. The first one is the antisymmetry of initial and final states in the calculation of matrix elements of the electromagnetic operators. It was found that the antisymmetry is crucial in M1 transitions in the region of resonances. At energies of astrophysical importance, the error introduced by neglecting the antisymmetry is however small. The second aspect is the role of the excited state $1/2^-$ of the target. At $E_{\rm CM} \sim 0$, the radiative capture cross sections in $^7$Be($p,\gamma$)$^8$B and $^7$Li($n,\gamma$)$^8$Li reactions are slightly impacted by the excited state of target. However, in the region of resonances and at higher energies the excited $1/2_1^-$ state in $^7$Be and $^7$Li turns out to be crucial. As compared to $^7$Be($p,\gamma$)$^8$B reaction, the $^7$Li($n,\gamma$)$^8$Li reaction is less sensitive to the first excited state of the target but more sensitive to the antisymmetry of initial and final states in the calculation of matrix elements of electromagnetic transition operators. The long wavelength approximation in the transition matrix elements changes mainly the E1 contribution to the radiative capture cross section.

\appendix

\section{Matrix elements of the electromagnetic transition operators}
\label{annexe_rad_cap_EM_operators}
The matrix elements related to electric and magnetic transitions will be considered with and without long-wavelength approximation. The operators involving the exact and approximate electromagnetic field
can be found in Ref. \cite{shalit04_b47}, whose matrix elements can be derived straightforwardly
from the Wigner-Eckhart theorem and standard manipulations of gradients of spherical harmonics coupled to angular momenta  \cite{varshalovich75_b12}. The operator ${ \hat{ \mathcal{M} }_{ L , {M}_{L} } }$ separates into electric ${ \hat{ \mathcal{M} }_{ L , {M}_{L} }^{E} }$ and magnetic ${ \hat{ \mathcal{M} }_{ L , {M}_{L} }^{M} }$ parts:
\begin{widetext}
	\begin{align}
		\hat{ \mathcal{M} }_{ L , {M}_{L} }^{E} &= \sum_{i} {e}_{i} \frac{ ( 2 L + 1 ) !! } { ( L + 1 ) {k}_{ \gamma }^{L} } \left[ \sphRicatiBessel{L}' ( {k}_{ \gamma } \hat{r}_{i} ) + \frac{ {k}_{ \gamma } \hat{r}_{i} }{2} \sphRicatiBessel{L} ( {k}_{ \gamma } \hat{r}_{i} ) \right] \hatY{L}{ {M}_{L} } ( { \Omega }_{i} ) \nonumber \\
		&+ \sum_{i} {e}_{i} \frac{ ( 2 L + 1 ) !! } { ( L + 1 ) {k}_{ \gamma }^{L} } \frac{ \hbar c }{ 2 {m}_{p} {c}^{2} } {g}^{s}_{i} \left[ \frac{ \sphRicatiBessel{L} ( {k}_{ \gamma } \hat{r}_{i} ) }{ \hat{r}_{i} } \right]  \left( \hatvec{l}_{i} \cdot \hatvec{s}_{i} \right) \hatY{L}{ {M}_{L} }( { \Omega }_{i} )
		\label{eq_rad_cap_ME_no_app}
	\end{align}
	\begin{align}
		\hat{ \mathcal{M} }_{ L , {M}_{L} }^{M} &= \frac{ \hbar c }{ 2 {m}_{p} {c}^{2} } \sum_{i} \frac{ ( 2 L + 1 ) !! }{ ( L + 1 ) {k}_{ \gamma }^{L} } \left[ {g}^{l}_{i} \vec{ \nabla }_{i} \left( \frac{ \sphRicatiBessel{L} ( {k}_{ \gamma } \hat{r}_{i} ) }{ {k}_{ \gamma } \hat{r}_{i} } \hatY{L}{ {M}_{L} } ( { \Omega }_{i} ) \right) \cdot \hatvec{l}_{i} + {g}^{s}_{i} \vec{ \nabla }_{i} \left( \sphRicatiBessel{L}' ( {k}_{ \gamma } \hat{r}_{i} ) \Y{L}{ {M}_{L} } ( { \Omega }_{i} ) \right) \cdot \hatvec{s}_{i} \right] \nonumber \\
		&+ \frac{ \hbar c }{ 2 {m}_{p} {c}^{2} } \sum_{i} \frac{ ( 2 L + 1 ) !! } { ( L + 1 ) {k}_{ \gamma }^{L} } {g}^{s}_{i} \left( {k}_{ \gamma } \sphRicatiBessel{L} ( {k}_{ \gamma } \hat{r}_{i} ) \right) \left( \hatvec{s}_{i} \cdot \uvec{ {r}_{i} } \right) \hatY{L}{ {M}_{L} } ( { \Omega }_{i} )  \nonumber \\
		&= - \frac{ \hbar c }{ 2 {m}_{p} {c}^{2} } \sum_{i} \frac{ ( 2 L + 1 ) !! }{ ( L + 1 ) {k}_{ \gamma }^{L} } {g}^{l}_{i} \sqrt{ \frac{ L + 1 }{ 2 L + 1 } } \left( \sphRicatiBessel{L}' ( {k}_{ \gamma } \hat{r}_{i} ) - \left( \frac{ L + 1 }{ \hat{r}_{i} } \right) \frac{ \sphRicatiBessel{L} ( {k}_{ \gamma } \hat{r}_{i} ) }{ {k}_{ \gamma } \hat{r}_{i} } \right) { [ \tensorop{Y}^{ L + 1 } ( { \Omega }_{i} ) \otimes \hatvec{l}_{i} ] }_{ {M}_{L} }^{L} \nonumber \\
		&+ \frac{ \hbar c }{ 2 {m}_{p} {c}^{2} } \sum_{i} \frac{ ( 2 L + 1 ) !! } { ( L + 1 ) {k}_{ \gamma }^{L} } {g}^{l}_{i} \sqrt{ \frac{L}{ 2 L + 1 } } \left( \sphRicatiBessel{L}' ( {k}_{ \gamma } \hat{r}_{i} ) + \left( \frac{L}{ \hat{r}_{i} } \right) \frac{ \sphRicatiBessel{L} ( {k}_{ \gamma } \hat{r}_{i} ) }{ {k}_{ \gamma } \hat{r}_{i} } \right) { [ \tensorop{Y}^{ L - 1 } ( { \Omega }_{i} ) \otimes \hatvec{l}_{i} ] }_{ {M}_{L} }^{L} \nonumber \\
		&- \frac{ \hbar c }{ 2 {m}_{p} {c}^{2} } \sum_{i} \frac{ ( 2 L + 1 ) !! }{ ( L + 1 ) {k}_{ \gamma }^{L} } {g}^{s}_{i} \sqrt{ \frac{ L + 1 }{ 2 L + 1 } } \left[ \left( \frac{ L ( L + 1 ) }{ { ( {k}_{ \gamma } \hat{r}_{i} ) }^{2} } - 1 \right) \sphRicatiBessel{L} ( {k}_{ \gamma } \hat{r}_{i} ) - \left( \frac{L}{ \hat{r}_{i} } \right) \sphRicatiBessel{L}' ( {k}_{ \gamma } \hat{r}_{i} ) \right] { [ \tensorop{Y}^{ L + 1 } ( { \Omega }_{i} ) \otimes \hatvec{s}_{i} ] }_{ {M}_{L} }^{L} \nonumber \\
		&+ \frac{ \hbar c }{ 2 {m}_{p} {c}^{2} } \sum_{i} \frac{ ( 2 L + 1 ) !! } { ( L + 1 ) {k}_{ \gamma }^{L} } {g}^{s}_{i} \sqrt{ \frac{L}{ 2 L + 1 } } \left[ \left( \frac{ L ( L + 1 ) }{ { ( {k}_{ \gamma } \hat{r}_{i} ) }^{2} } - 1 \right) \sphRicatiBessel{L} ( {k}_{ \gamma } \hat{r}_{i} ) + \left( \frac{ L + 1 }{ \hat{r}_{i} } \right) \sphRicatiBessel{L}' ( {k}_{ \gamma } \hat{r}_{i} ) \right] { [ \tensorop{Y}^{ L - 1 } ( { \Omega }_{i} ) \otimes \hatvec{s}_{i} ] }_{ {M}_{L} }^{L} \nonumber \\
		&+ \frac{ \hbar c }{ 2 {m}_{p} {c}^{2} } \sum_{i} \frac{ ( 2 L + 1 ) !! }{ ( L + 1 ) {k}_{ \gamma }^{L} } {g}^{s}_{i} \left( {k}_{ \gamma } \sphRicatiBessel{L} ( {k}_{ \gamma } \hat{r}_{i} ) \right) \left( \hatvec{s}_{i} \cdot \uvec{ {r}_{i} } \right) \hatY{L}{ {M}_{L} } ( { \Omega }_{i} )
		\label{eq_rad_cap_MM_no_app}
	\end{align}
\end{widetext}
where
${ i }$ runs over all considered nucleons and ${ \hatvec{l}_{i} }$ and ${ \hatvec{s}_{i} }$ are the orbital and spin angular momenta, respectively. In the above expression, 
${ \hatY{L}{ {M}_{L} } ( \Omega ) }$ is a spherical harmonics,
${ \sphRicatiBessel{L} }$ is the Ricatti-Bessel function,
${ \hat{r}_{i} }$, ${ { \Omega }_{i} }$ are radial and angular coordinates of the nucleon ${ i }$, and ${ \uvec{ {r}_{i} } = \hatvec{r}_{i} / \hat{r}_{i} }$. Moreover,
${ {e}_{i} }$ is the dimensionless charge of the nucleon ${ i }$ (${ {e}_{i} }=1$ for a proton and 0 for a neutron),
${ {g}^{s}_{i} }$ is the dimensionless magnetic spin moment of the nucleon ${ i }$ (${ {g}^{s}_{i} }=5.5857$ for a proton and -3.8263 for a neutron), 
${ {m}_{p} {c}^{2} }$ (in units of MeV) is the mass of the proton, and 
${ {g}^{l}_{i} }$ is the dimensionless magnetic orbital momentum of the nucleon ${ i }$ times ${ L + 1 }$ (${ {g}^{l}_{i} }=2$ for a proton and 0 for a neutron).

In the long wavelength approximation, the expressions \eqref{eq_rad_cap_ME_no_app} and \eqref{eq_rad_cap_MM_no_app} become:
\begin{equation}
	\hat{ \mathcal{M} }_{ L , {M}_{L} }^{E} = \sum_{i} {e}_{i} \hat{r}_{i}^{L} \hatY{L}{ {M}_{L} } ( { \Omega }_{i} )
	\label{eq_rad_cap_ME_app}
\end{equation}
\begin{widetext}
	\begin{align}
		\hat{ \mathcal{M} }_{ L , {M}_{L} }^{M} &= \frac{ \hbar c }{ 2 {m}_{p} {c}^{2} } \sum_{i} \left[ \frac{ {g}^{l}_{i} }{ L + 1 } \vec{ \nabla }_{i} \left( \hat{r}_{i}^{L} \hatY{L}{ {M}_{L} } ( { \Omega }_{i} ) \right) \cdot \hatvec{l}_{i} + {g}^{s}_{i} \vec{ \nabla }_{i} \left( {r}^{L}_{i} \Y{L}{ {M}_{L} } ( { \Omega }_{i} ) \right) \cdot \hatvec{s}_{i} \right] \nonumber \\
		&= \frac{ \hbar c }{ 2 {m}_{p} {c}^{2} } \sum_{i} \sqrt{ L ( 2 L + 1 ) } \hat{r}_{i}^{ L - 1 } \frac{ {g}^{l}_{i} }{ L + 1 } { [ \tensorop{Y}^{ L - 1 } ( { \Omega }_{i} ) \otimes \hatvec{l}_{i} ] }^{ {L}_{ {M}_{L} } } \nonumber \\
		&+ \frac{ \hbar c }{ 2 {m}_{p} {c}^{2} } \sum_{i} \sqrt{ L ( 2 L + 1 ) } \hat{r}_{i}^{ L - 1 } {g}^{s}_{i} { [ \tensorop{Y}^{ L - 1 } ( { \Omega }_{i} ) \otimes \hatvec{s}_{i} ] }^{ {L}_{ {M}_{L} } }
		\label{eq_rad_cap_MM_app}
	\end{align}
\end{widetext}
Eqs. \eqref{eq_rad_cap_ME_no_app}-\eqref{eq_rad_cap_MM_app} have been written in such a way that only one-body operators appear in each summation. Matrix elements of these operators are calculated in a standard way using the Wigner-Eckhart theorem. 

\begin{acknowledgments}
	This work was supported partially through FUSTIPEN (French-U.S. Theory Institute for Physics with Exotic Nuclei) under DOE grant number DE-FG02-10ER41700 and by the DOE award number DE-FG02-96ER40963 (University of Tennessee). One of the authors (M.P.) wish to thank COPIN and COPIGAL for the support.	
\end{acknowledgments}

\bibliographystyle{apsrev4-1}
\bibliography{apsrev_refer,book,thesis}

\end{document}